\documentclass[twocolumn, twocolappendix]{aastex631}

\usepackage{multirow}
\usepackage{natbib}

\newcommand{\Lya}{Ly$\alpha$}

\newcommand{\Hb}{H$\beta$}
\newcommand{\Ha}{H$\alpha$}
\newcommand{\MgII}{Mg\,\textsc{II}}
\newcommand{\OIII}{[O\,\textsc{III}]}

\newcommand{\CIV}{C\,\textsc{IV}}

\newcommand{\ergs}{\text{erg s$^{-1}$}}

\newcommand{\kms}{\text{km s$^{-1}$}}
\newcommand{\rpi}{{\rm \pi}}

\hypersetup{
   colorlinks,
   linkcolor={blue!88!black!80},
   citecolor={blue!88!black!80},
   urlcolor={blue!88!black!80}}
   
\received{August 27, 2021}
\revised{November 6, 2021}
\accepted{November 8, 2021}
\submitjournal{ApJ}

\shorttitle{Composite spectra of EVQs}
\shortauthors{Ren W. et al.}

\begin{document}

\title{Extreme Variability Quasars in Their Various States. I: The sample selection and composite SDSS spectra}

\correspondingauthor{Wenke Ren, Junxian Wang, Zhenyi Cai}
\email{rwk@mail.ustc.edu.cn, jxw@ustc.edu.cn, zcai@ustc.edu.cn}

\author[0000-0002-3742-6609]{Wenke Ren}
\affiliation{CAS Key Laboratory for Research in Galaxies and Cosmology, Department of Astronomy, University of Science and Technology of China, Hefei, Anhui 230026, China}
\affiliation{School of Astronomy and Space Science, University of Science and Technology of China, Hefei 230026, China}

\author[0000-0002-4419-6434]{Junxian Wang}
\affiliation{CAS Key Laboratory for Research in Galaxies and Cosmology, Department of Astronomy, University of Science and Technology of China, Hefei, Anhui 230026, China}
\affiliation{School of Astronomy and Space Science, University of Science and Technology of China, Hefei 230026, China}

\author[0000-0002-4223-2198]{Zhenyi Cai}
\affiliation{CAS Key Laboratory for Research in Galaxies and Cosmology, Department of Astronomy, University of Science and Technology of China, Hefei, Anhui 230026, China}
\affiliation{School of Astronomy and Space Science, University of Science and Technology of China, Hefei 230026, China}

\author[0000-0001-8416-7059]{Hengxiao Guo}
\affiliation{Department of Physics and Astronomy, 4129 Frederick Reines Hall, University of California, Irvine, CA, 92697-4575, USA}

\begin{abstract}
   Extremely variable quasars (EVQs) are a population of sources showing large optical photometric variability revealed by time-domain surveys.  
   The physical origin of such extreme variability is yet unclear.
   In this first paper of a series, we construct the largest-ever sample of 14,012 EVQs using photometric data spanning over $>$ 15 years from SDSS and Pan-STARRS1. 
   We divide them into five sub-samples according to the relative brightness of an EVQ during SDSS spectroscopic observation compared to the mean brightness from photometric observations.
   Corresponding control samples of normal quasars are built with matched redshift, bolometric luminosity and supermassive black hole mass.
   We obtain the composite SDSS spectra of EVQs in various states and their corresponding control samples.
   We find EVQs exhibit clearly bluer (redder) SDSS spectra during bright (dim) states, consistent with the ``bluer-when-brighter" trend widely seen in normal quasars.
   We further find the line EWs of broad \MgII, \CIV\ and \OIII\ (but not broad \Hb\ which is yet puzzling) gradually decreases from dim state to bright state, similar to the so-called intrinsic Baldwin effect commonly seen in normal AGNs.
   Meanwhile, EVQs have systematically larger line EWs compared with the control samples. 
   We also see that EVQs exhibit subtle excess in the very broad line component compared with control samples. 
   Possible explanations for the discoveries are discussed. Our findings support the hypothesis that EVQs are in the tail of a broad distribution of quasar properties, but not a distinct population.
\end{abstract}


\section{Introduction} \label{sec:intro}

Quasars and active galactic nuclei (AGNs) can be observationally classified into type 1 and 2, i.e., with and without broad emission lines (BELs), respectively. 
The unified model indicates that two types of objects are intrinsically identical but merely viewed at different orientations \citep{Antonucci1993,Urry1995c}. 
With the cumulus of long-term repeated observations, a unique and rare class of quasars showing dramatic emergence or disappearance of the broad emission lines has been discovered \citep[e.g.,][]{Denney2014b,LaMassa2015a,Runnoe2015,Ruan2015,MacLeod2016,McElroy2016a,Stern2018,Wang2018,Yang2018,MacLeod2019,Trakhtenbrot2019b,Sheng2020}.
These objects, dubbed as ``changing-look'' quasars (CLQs), often also show strong UV/optical continuum variations with a factor $\gtrsim$ 10 \citep[e.g.,][]{MacLeod2016} within a typical timescale of decades.
The prominent changes in the BELs and continuum on such short timescales are difficult to be explained by the traditional thin disk theory which predicts a much longer timescale (associated with the viscous timescale, typically $\sim 10^4 yr$) for accretion rate change. 
Instead they are more likely associated with changes of unclear cause in the innermost regions of the accretion disk \citep[e.g.,][]{Ross2018,Stern2018}.

The variability is a hallmark signature of quasars across all wavelength and timescale \citep[e.g.,][]{Mushotzky1993,Ulrich1997}.
In rest-frame UV/optical, the continuum emission from the accretion disc of quasars typically varies by $\sim0.2$ mag on timescales of days to years \citep[e.g.,][]{VandenBerk2004,Sesar2007}.
However, it is yet unclear whether such normal variations in quasars and the rare strong variations in CLQs are caused by the same mechanism(s). 

Recently \cite{Rumbaugh2018} identified $\sim$ 1000 quasars with extreme variability (EVQs) in $g$-band ($\Delta g> 1$ mag, $\sim$ 10\% of all quasars searched) utilizing photometric light curves spanning more than 15 years. 
\cite{Rumbaugh2018} found EVQs have larger BEL equivalent widths (EW) and lower Eddington ratio compared with a control sample with matched redshift and optical luminosity, and suggested
``EVQs seem to be in the tail of a continuous distribution of quasar properties, rather than standing out as a distinct population."
Strikingly, \cite{MacLeod2019} spectroscopically confirmed $\gtrsim$ 20\% of EVQs they selected as CLQs, i.e., CLQs belong to a subset of EVQs.
EVQs, which are more common than CLQs and can be identified merely with photometry, 
are thus ideal targets of statistical studies \citep[e.g.][]{Rumbaugh2018,Luo2020}. 
It is thus worth to build larger samples of EVQs and to explore whether they have other special physical properties different from ordinary quasars.
Meanwhile, multiple spectroscopic observations, which are essential to probe the spectral variabilities in EVQs, have only been reported for very small samples \citep[][]{MacLeod2019,Yang2020,Guo2020}.

In this work, we compile a large sample of 14,012 EVQs selected using SDSS and Pan-STARRS1 photometric observations of SDSS quasars. 
We point out that, for an EVQ with only single epoch spectroscopy available, comparing its synthetic spectroscopy magnitude with mean magnitude from the photometric light curve, one can indeed identify whether the spectrum was obtained during a bright, normal or dim state.
We thus could divide the sample into five subsamples according to the deviation of SDSS synthetic spectroscopy magnitude from the mean photometric magnitude, i.e., EVQs caught in their
extremely bright state, bright state, median state, dim state and extremely dim state, respectively. 
We study the spectral properties (line EWs, co-added spectra and line profiles) and their evolution (from dim to bright states) of EVQs through comparing the five subsamples with corresponding control samples with matched redshift, luminosity, and supermassive black hole (SMBH) mass.

We stress that comparison of EVQs with control samples with matched luminosity and SMBH mass (thus matched Eddington ratio) is essential to this work, as the spectral properties of quasars could be sensitive to these parameters. For instance, while it is well known that the UV/optical emission line EW is anti-correlated with the underlying continuum luminosity \citep[the so-called ensemble Baldwin effect, eBeff, e.g.][]{Baldwin1977,Dietrich2002},
the Eddington ratio could be one of the dominant key factors behind the eBeff \citep{Baskin2004,Dong2009}.

Another closely relevant phenomenon is the ``intrinsic Baldwin effect" (iBeff), that for individual AGNs, it has also been found that the line EW decreases when AGNs brighten \citep[e.g.][]{Pogge1992,Kinney1990,Homan2020}, 
and the slope of the iBeff is usually steeper than the eBeff for the same lines \citep{Pogge1992,Kinney1990}.
Comparing the spectra of EVQs among different brightness states, we would be able to examine whether EVQs also exhibit intrinsic Baldwin effect. 

The structure of this paper is as follows.
We describe the data and reduction in \S\ref{sec:data}.
The selection criteria for EVQs and their control samples are presented in \S\ref{sec:sample}.
In \S\ref{sec:results} we present the composite spectra for EVQs (and control samples) and their emission line properties. 
We discuss our results in \S\ref{sec:discussion} and summarize in \S\ref{sec:conclusion}.
Throughout this paper we adopt a flat $\rm \Lambda CDM$ cosmology with $\Omega _\Lambda = 0.7$, $\Omega _{m} = 0.3$ and $H_0 = 70\ {\rm km\ s^{-1}\ Mpc^{-1}}$.

\section{The Data and Reduction} \label{sec:data}

\subsection{Photometric Observations}\label{subsec:photometry}

We start from the Sloan Digital Sky Surveys (SDSS) date release 14 (DR14) quasar catalog \citep[DR14Q,][]{Paris2018}, which contains 526,356 spectroscopically confirmed quasars with luminosity $M_i[z=2]<-20.5$ over 9376 $\rm deg^2$. 
Note the DR14Q catalog only provides single epoch photometry (i.e., the primary SDSS magnitude) for each source. In order to select EVQs, we construct the $g$- and $r$-band light curves for each quasar through further gathering all archive photometric observations from both SDSS\footnote{\url{https://skyserver.sdss.org/casjobs/}} and Pan-STARRS1\footnote{\url{https://mastweb.stsci.edu/ps1casjobs/home.aspx}} (PS1) databases. 

The SDSS photometric observations were taken by the drift-scan camera (30 2k $\times$ 2k CCDs) installed on the 2.5m Sloan telescope \citep{Gunn2006}. 
In the SDSS DR14Q catalog, the primary photometric observations were obtained with SDSS-I/II, lasting from 2000 to 2007 (covering 11,663 $\rm deg^{2}$), and SDSS-III (before 2009) on $\rm \sim3000\ deg^{2}$ new sky area.
Using a matching radius of 1\arcsec, we gather all available SDSS $g$-and $r$-band photometry for DR14Q quasars.
Referring to the recommendation on the use of photometric processing flags from SDSS\footnote{\url{https://www.sdss.org/dr16/algorithms/photo_flags_recommend/}}, we reserve detections with $\rm{mode=1}$ or 2, ``clean'' flags and PSF magnitude error $<$ 0.2 mag in both $g$- and $r$-band.
About 45\% quasars (237,099 quasars, specifically) have multi-epoch SDSS photometry (and 28,672 of them had been observed over more than five epochs).

We also collect the $g$- and $r$-band photometry from the PS1 $3\rpi$ survey, with up to four exposures in each band per year, conducted from 2009 to 2013.
In total, within a matching radius of 1\arcsec, 510,838 SDSS DR14 quasars have counterparts in the PS1 archive (with mean epochs of around 10 and 12 in $g$- and $r$-band, respectively).
Similar to the processing of SDSS data, we also filter the matched PS1 detections according to their photometric info flags.
We rule out PS1 detections with PSF magnitude error $>$ 0.2 mag or flagged as:
\begin{enumerate}
   \item Peak lands on diffraction spike, ghost or glint;
   \item Poor moments for small radius, try large radius or could not measure the moments;
   \item Source fit failed or succeeds but with low S/N, high Chi square or too large for PSF;
   \item Source model peak is above saturation;
   \item Size could not be determined;
   \item Source has crNsigma above limit;
   \item Source is thought to be a defect;
   \item Failed to get good estimate of object's PSF;
   \item Detection is astrometry outlier.
\end{enumerate}
After conversion between PS1 and SDSS photometry (see \S\ref{subsec:conversion}), we could merge photometric data points from both SDSS and PS1, and obtain for every quasar relatively long-term $g-$ \& $r-$ band light curve with a length of $4 \sim 15$ years. 
We further remove from the light curves a small fraction of photometric measurements which could be unreliable according to the consistency check between  $g$- and $r$-band light curves (see \S\ref{subsec:rejectbad}). 

A mean magnitude $g_{\rm mean}$ is then derived from the clean light curve for each quasar. We need such a mean magnitude to represent the average brightness of a quasar over a long duration, and to be compared with the synthetic magnitude $g_{\rm spec}$ derived from the SDSS  spectrum to determine if the spectrum was captured during a bright, normal or dim state. In order to avoid the mean magnitude being over-dominated by a few measurements with very small uncertainties, we simply calculate the mean without error weighting.  
The light curves of most quasars contain far more data points from PS1 than from SDSS (and contrarily for quasars in SDSS Stripe 82). In order to avoid the mean magnitudes being over-dominated by data from one instrument or by data from a single observing season with a large number of epochs, we first calculate the yearly mean magnitudes and then the final mean from the yearly mean values.

\subsection{Photometric Magnitude Conversion} \label{subsec:conversion}

\begin{figure}[tb!]
   \includegraphics[width=.48\textwidth]{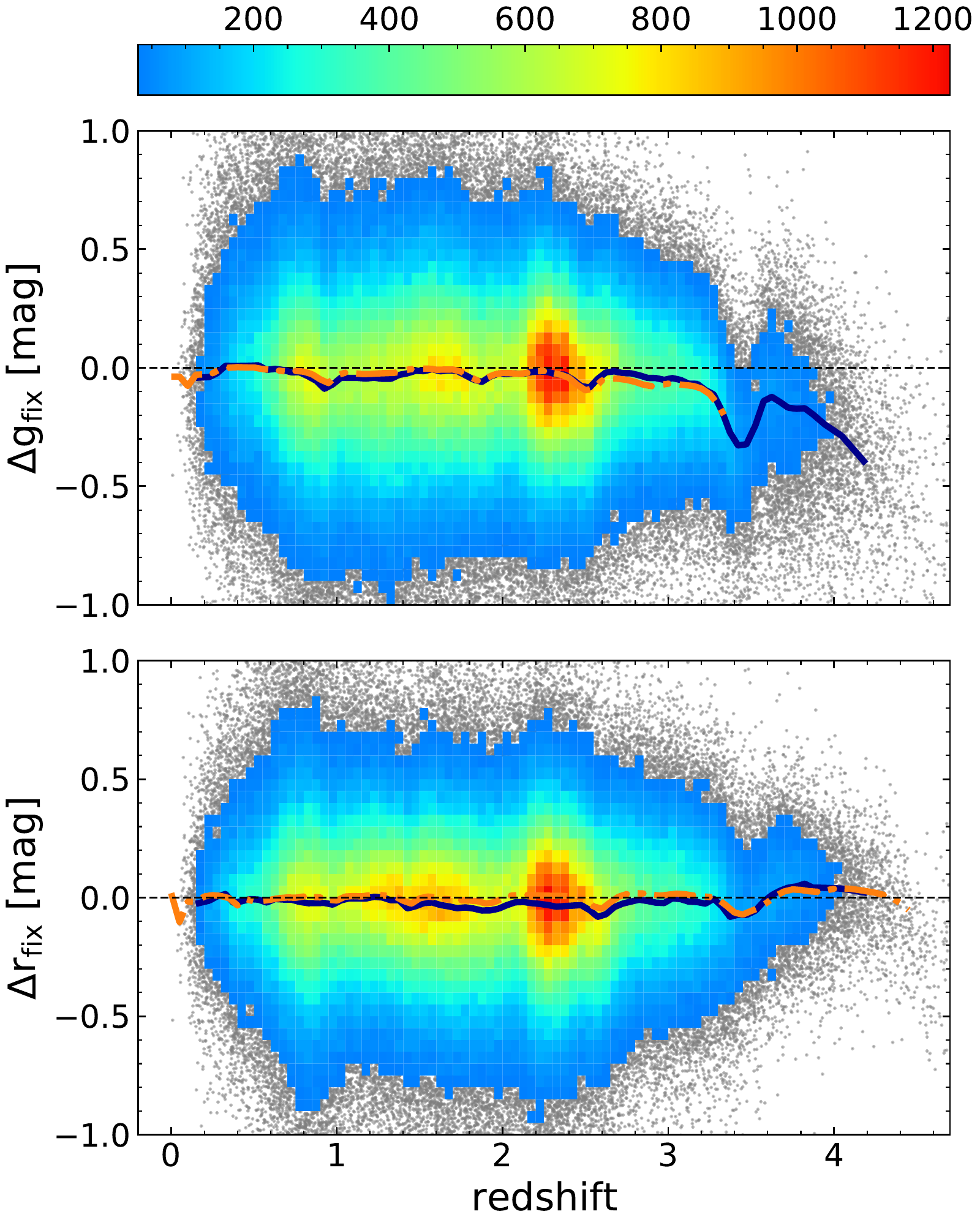}
   \caption{The distributions of photometry difference between PS1 and SDSS, $\Delta g_{\rm fix} = g_{\rm PS1} - g_{\rm SDSS}$ (top panel) and $\Delta r_{\rm fix} = r_{\rm PS1} - r_{\rm SDSS}$ (bottom panel), versus redshift for SDSS DR14 quasars. In each panel, the colored contour represents the distribution density, while the dark-blue line represents the mean photometry difference in redshift bin of 0.05. For comparison, assuming the mean quasar spectrum from \cite{Yip2004}, the $g$- and $r$-band filter differences between PS1 and SDSS are shown as the orange dot-dashed lines.
   \label{fig:magfix}}
\end{figure}

In Fig. \ref{fig:magfix} we plot the distributions of $\Delta g_{\rm fix}$ =$g_{\rm PS1} - g_{\rm SDSS}$  (and $\Delta r_{\rm fix}$ =$r_{\rm PS1} - r_{\rm SDSS}$) versus redshift for SDSS-PS1 matched quasars, where 
$g_{\rm PS1}$ ($r_{\rm PS1}$) and $g_{\rm SDSS}$ ($r_{\rm SDSS}$) are the mean PS1 magnitude and the primary SDSS magnitude, respectively.
Subtle but clear offsets between PS1 and SDSS magnitudes (see the dark-blue lines which plot the mean offsets between PS1 and SDSS photometries averaged with a redshift bin of $\Delta z = 0.05$) are seen within certain redshift ranges. This is owing to the slight difference of the filter transmission curves between PS1 and SDSS.
For example, given that the red side transmission of the PS1 $g$-band is considerably higher than that of SDSS $g$ at $5200 \sim 5500$\AA, where locates the \MgII\ emission line for quasars at $z \sim 0.9$, PS1 $g$-band would receive more \MgII\ emission line photons and result in a brighter magnitude compared with SDSS $g$. 
Moreover, the general decreasing of the $\Delta g_{\rm fix}$ beyond $z \sim 2.6$ could be due to the fact that the spectral slope of quasars is redder at the wavelength shorter than 1216\AA\ (intrinsically and/or because of intergalactic medium absorption), which makes the energy distribution of the photons that $g$-band received concentrates to the red side and magnifies the effect of filter transmission difference near $5200 \sim 5500$~\AA\ we mentioned before.
The dependence on the redshift of the mean offset could indeed be well recovered through comparing the synthetic PS1 and SDSS magnitudes of the mean quasar spectrum of  \cite{Yip2004} (the dot-dashed lines in Fig. \ref{fig:magfix}). Note the mean spectrum extends down to 900~\AA~ in the rest frame thus results in a cut at $z \sim 3.4$ (for $g$ band).

To eliminate the systematic offsets between PS1 and SDSS magnitudes for our quasars, a correction is applied to the PS1 magnitudes of each quasar using the mean $\Delta g_{\rm fix}$ ($\Delta r_{\rm fix}$) of its closest 1000 neighbors in redshift space.

\subsection{Reject defects in photometry} \label{subsec:rejectbad}
\begin{figure}[tb!]
   \includegraphics[width=.48\textwidth]{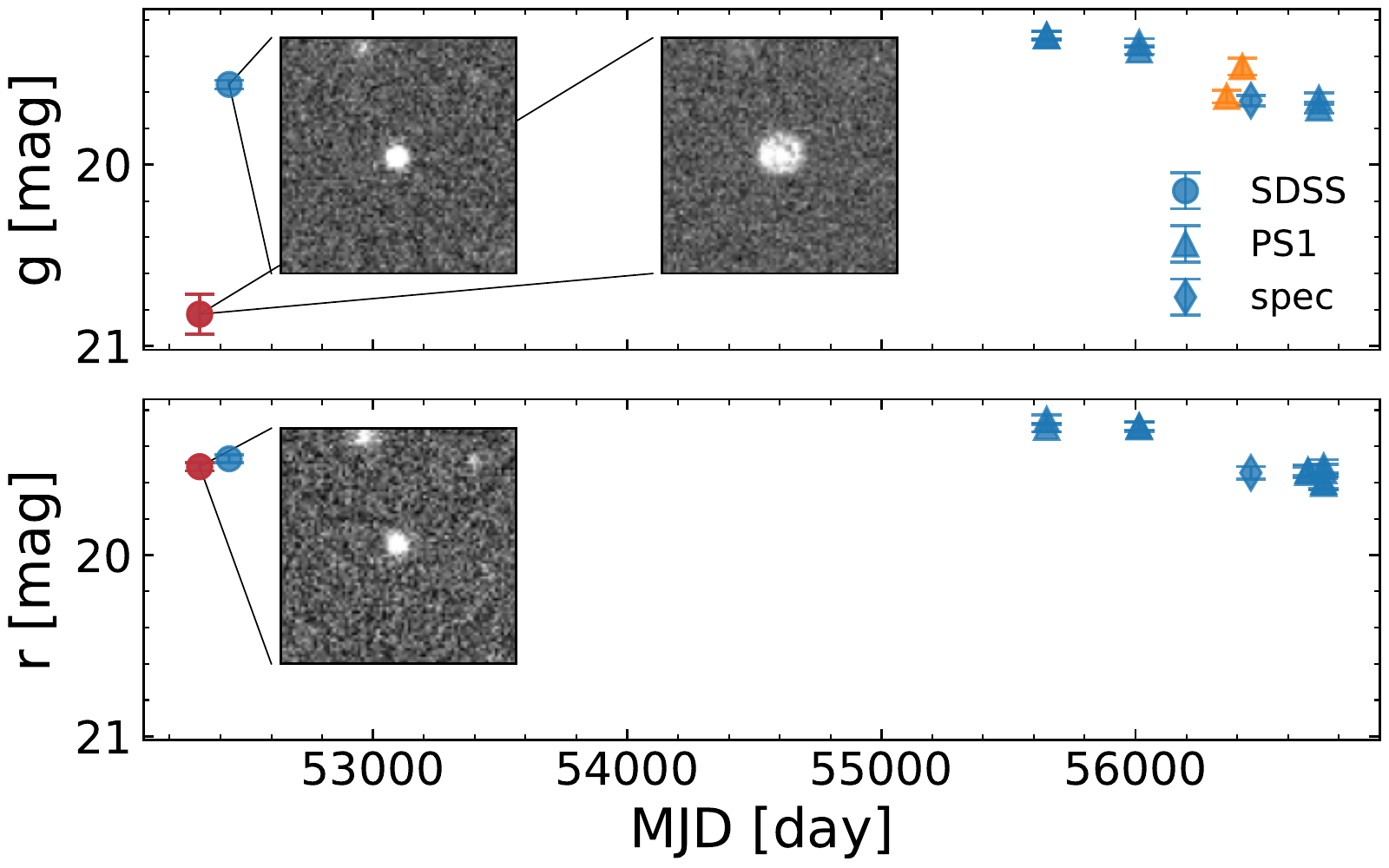}
   \caption{The $g$- (top) and $r$-band (bottom) light curve for SDSS J135036.05+584853.8. 
   The detections at ${\rm MJD}=52318$, which give an extraordinarily dim $g$-band magnitude while with a moderate $r$-band magnitude, are marked in red. The image cutout with 30\arcsec $\times$ 30\arcsec\ for both bands are given, showing the $g$-band photometry is problematic. As comparison, another good $g$-band image 116 days later is given.
   The synthetic $g$- and $r$- band magnitudes derived from its SDSS spectrum are also plotted.
   The yellow data points mark epochs we have excluded in \S\ref{subsec:photometry}.
   \label{fig:defectSDSS}}
\end{figure}
\begin{figure}[tb!]
   \includegraphics[width=.48\textwidth]{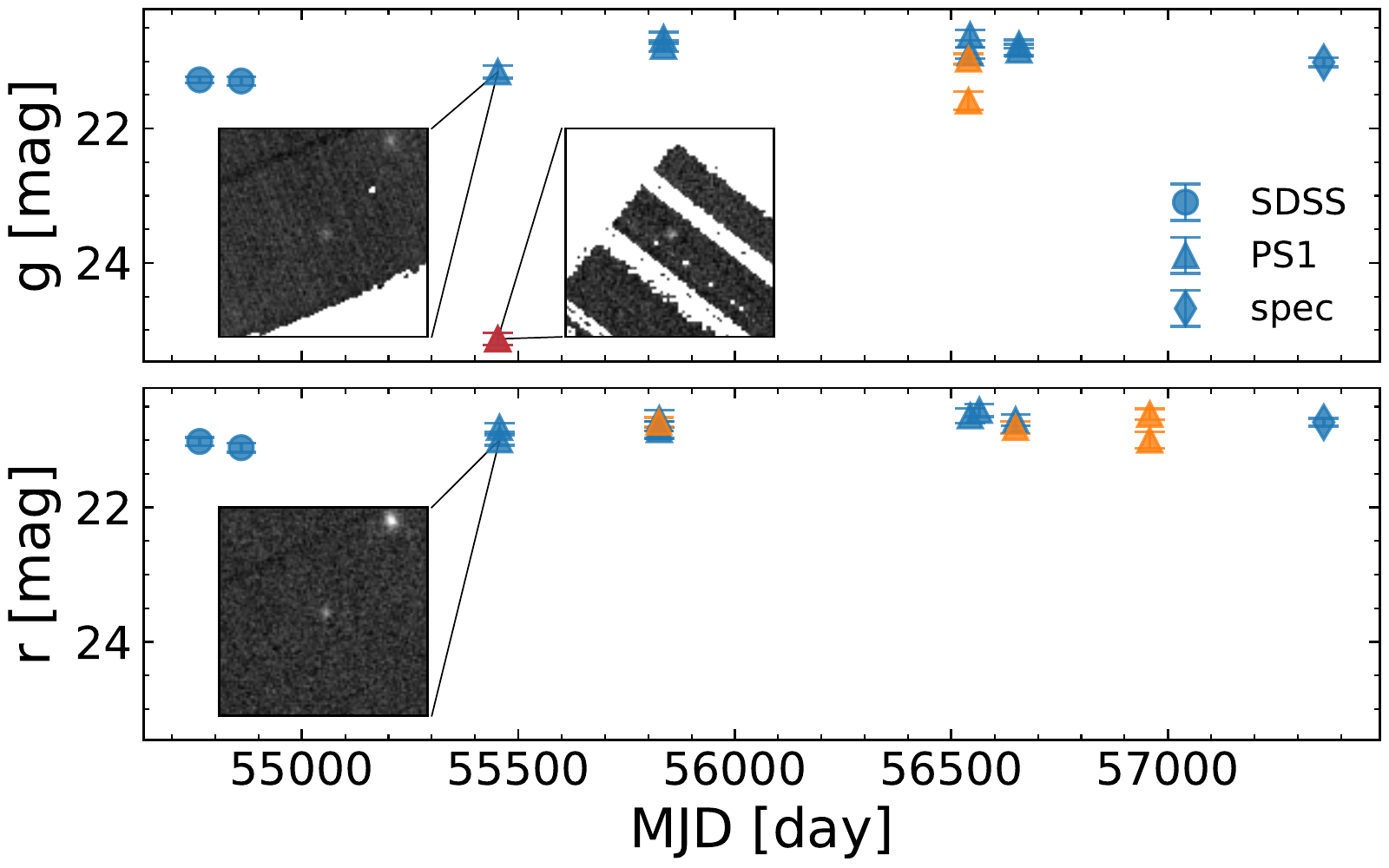}
   \caption{The $g$- (top) and $r$-band (bottom) light curve for SDSS J004134.29+282844.4. Symbols are the same as in Fig. \ref{fig:defectSDSS}. The 30\arcsec $\times$ 30\arcsec\ image cutout of the defect PS1 detection in $g$-band (marked in red) is over-plotted. The cutouts of nearest good epochs in $g$- and $r$-band are over-plotted for comparison.
   \label{fig:defectPS1}}
\end{figure}
\begin{figure}[tb!]
  \includegraphics[width=.48\textwidth]{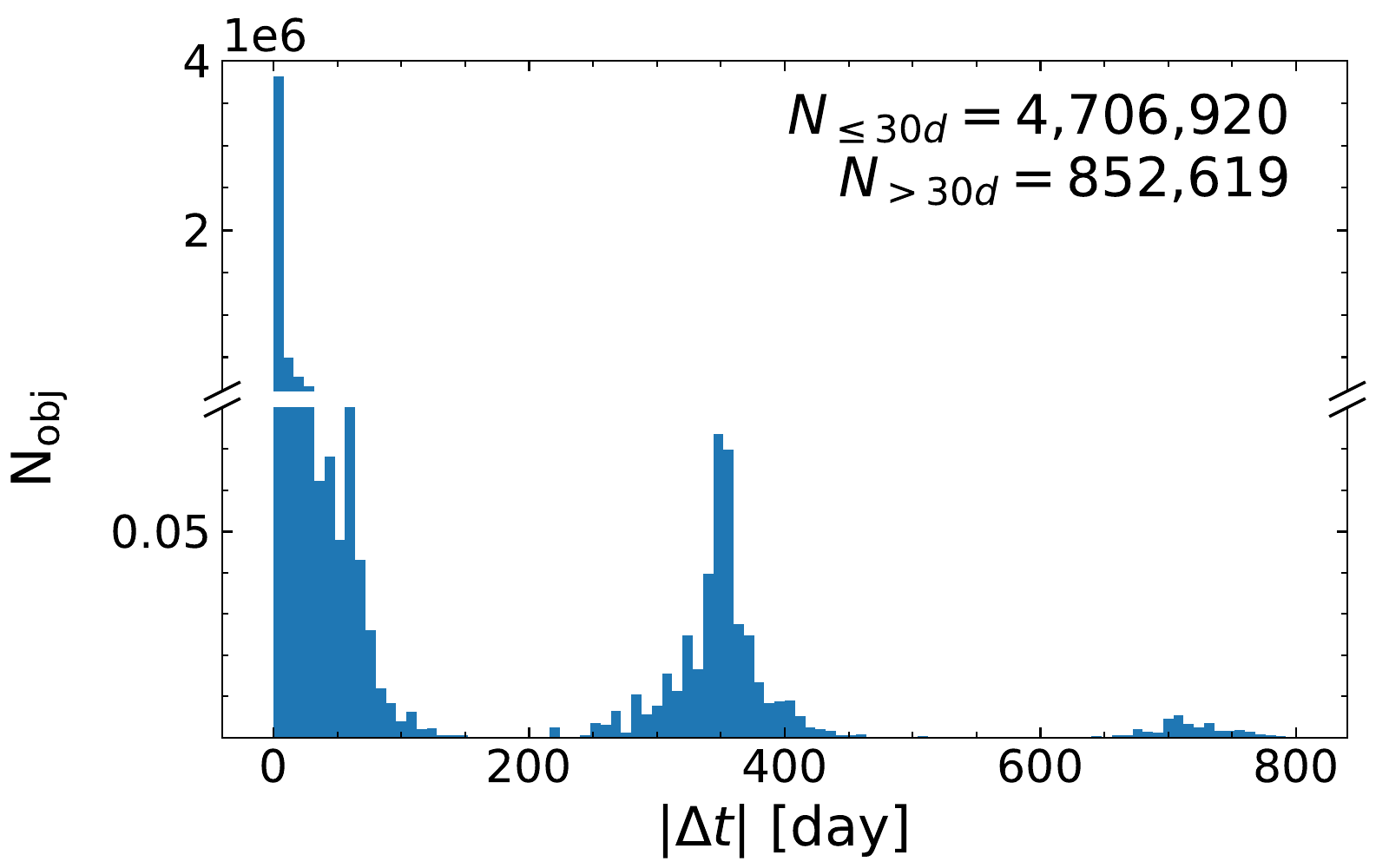}
  \caption{The distribution of the interval between each PS1 $g$-band exposure and its nearest in time $r$-band PS1 counterpart. 
  \label{fig:epochinterval}}
\end{figure}
\begin{figure}[tb!]
  \includegraphics[width=.48\textwidth]{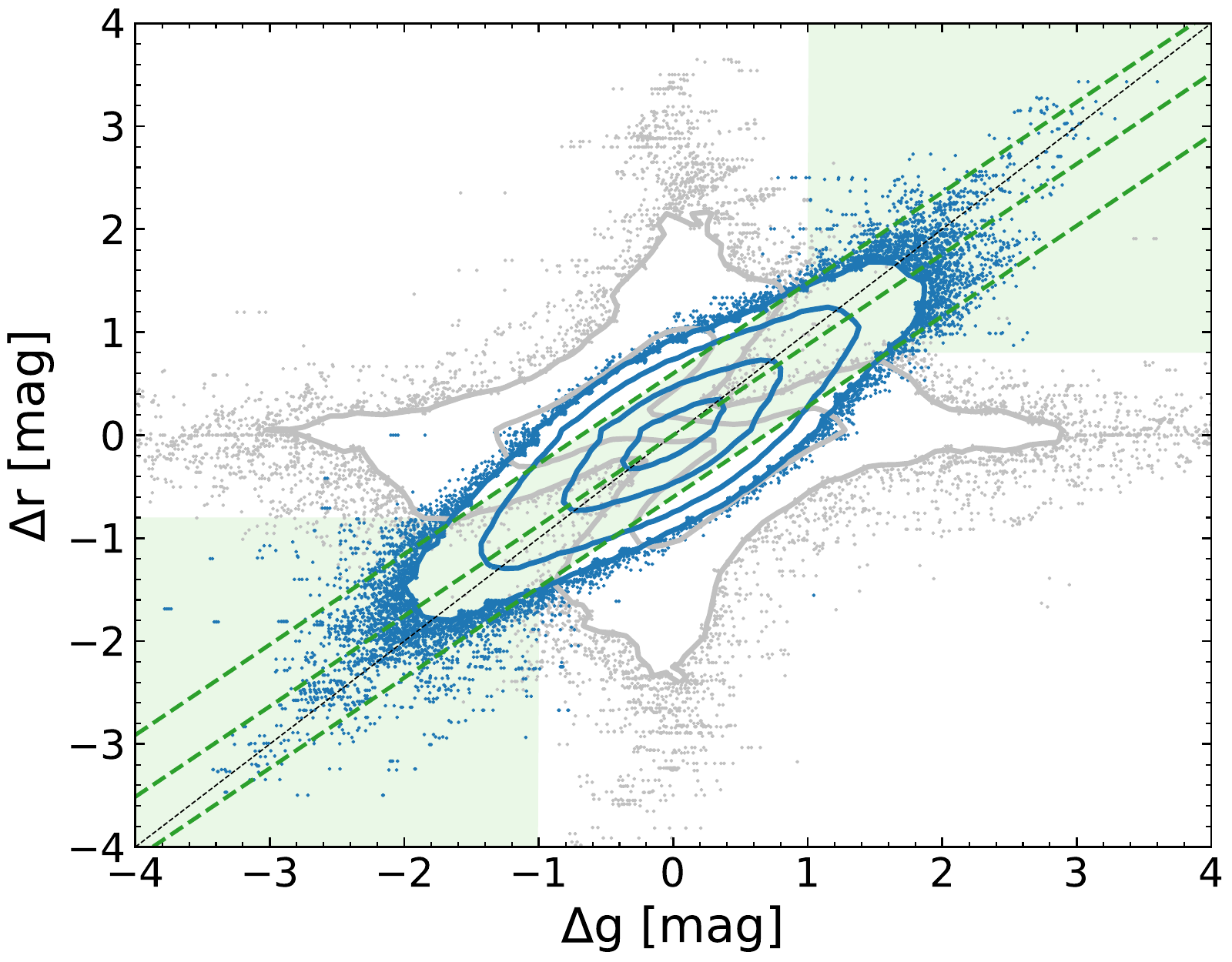}
  \caption{
  The $g$- and $r$-band magnitude variability between any pairs of epochs. Epoch pairs from all quasars are plotted. 
  The green dashed line is the orthogonal distance regression (ODR) fitting result of the whole dataset with a linear model. We define a green zone within which $\Delta g$ and $\Delta r$ of a pair of epochs appear consistent with each other, i.e., is unlikely affected by photometry defect (see text for details). 
  Aided by the green zone, we identify ``problematic" epochs potentially affected by photometry defects (either in $g$ or $r$ band). The epoch pairs with ``problematic" epochs involved, accounting for 0.49\% of the whole dataset, are plotted in gray (with $1\sigma$ and $2\sigma$ density contours).
  The blue dots (with $1\sigma$, $2\sigma$, $3\sigma$ and 99.97\% density contours) are free from ``problematic" epochs, representing 99.51\% of the whole data set.
  \label{fig:detectiondiff}}
\end{figure}

A commonly used criterion for selecting EVQs is to ask for a change in magnitude of $|\Delta g|>1$ mag between any two epochs in the light curve \citep[e.g.,][]{MacLeod2016, Rumbaugh2018,Guo2020}.
By applying $|\Delta g|>1$ on our sample, we can get $\sim$ 56,000 candidates, however over 1/6 of them have $|\Delta r|<0.5$ which are highly suspected and are likely due to defects in photometry measurements, such as by ghost, glint, cosmic ray, CCD problem or unknown instrumental problems. 
Such defects should be rejected before we could build a reliable sample of EVQs.

In Fig. \ref{fig:defectSDSS} and Fig. \ref{fig:defectPS1}, we present for example the $g$- and $r$-light curves for two sources, each showing a clear defect in $g$-band photometry during an individual epoch.
In Fig. \ref{fig:defectSDSS} the SDSS $g$-band cutout of the problematic epoch (marked in red) shows much blurry and diffuse signal comparing with the corresponding $r$-band data and another $g$-band image obtained 116 days later.
In Fig. \ref{fig:defectPS1} the red dot mark an epoch during which the PS1 $g$-band signal is clearly contaminated by an artificial CCD feature.

Such epochs could be identified through checking the consistency between $g$- and $r$-band variability revealed in the light curves. 
For SDSS, simultaneous $g$- and $r$-band photometry are always available.
However, this is not true for PS1.
For each PS1 $g$-band photometry we identify an $r$-band counterpart which was obtained closest in time. 
In Fig. \ref{fig:epochinterval} we plot the distribution of the time intervals between $g$-band and the corresponding (closest in time) $r$-band exposures.
For most PS1 $g$-band exposure we could pair it with an $r$-band exposure obtained within 30 days, and use this quasi-simultaneous $r$-band data to examine the reliability of the $g$-band photometry (assuming a quasar does not significantly vary within a month). We simply drop those PS1 $g$-band exposures without corresponding $r$-band exposure obtained within 30 days.

Then for an individual quasar, we could calculate $\Delta g$ and $\Delta r$ (a later measurement minus an earlier measurement) between any pair of epochs, and we plot $\Delta g$ vs $\Delta r$ for all quasars in Fig. \ref{fig:detectiondiff}.
We expect that the variability revealed with good photometry should follow 
the general trend seen in the $\Delta g$ and $\Delta r$ plot (orthogonal distance regression yields  $\Delta r = 0.878 \Delta g$). 
In Fig. \ref{fig:detectiondiff} we define a green zone within which the variability well follows the general trend: 
\begin{eqnarray}
   |\Delta r - 0.878\Delta g|<0.6\\
   \Delta g <-1 ~and~ \Delta r < -0.8\\
   \Delta g >1 ~and~ \Delta r >0.8
\end{eqnarray}
and consider those pairs out of the green zone potentially affected by defects in photometry. 
For a single quasar with N epochs in the light curve, an individual epoch contributes to N$-$1 pairs of epochs. If more than a half of the N$-$1 pairs of epochs locate out of the green zone, we mark this epoch problematic and then drop it from the light curves. 
A total of 14,899 epochs (0.27\% out of 5,559,539) are marked ``problematic" with this approach and excluded.  
In Fig. \ref{fig:detectiondiff}, pairs with the ``problematic" epoch involved are plotted in gray, yielding a butterfly-shaped distribution and demonstrating the inconsistency between $\Delta g$ and $\Delta r$ caused by photometry defects.

\subsection{Spectroscopic Observations and Spectral Fitting} \label{subsec:specfit}

We consider all available SDSS spectra for the DR14Q quasars and compare their synthetic magnitudes (spectrophotometry) of the spectra with the aforementioned photometric mean magnitudes (see \S\ref{subsec:photometry}) to determine their spectral states.
Same as the photometric magnitude, we demand the g-band synthetic magnitude error $<$ 0.2 mag.
In total, about 1/7 of the spectra were taken during SDSS-III or earlier, about a half taken after PS1.

We fit the quasar spectra mainly following \citet[][ hereafter S11]{Shen2011} and using the {\tt PyQSOFit} code \citep{2018ascl.soft09008G}.
For each spectrum, after correcting for the Galactic extinction adopting the dust map of \cite{Schlegel1998} and the \cite{Fitzpatrick1999} extinction law assuming ${\rm RV}=3.1$, we shift the spectrum to the rest frame using the redshift given in the DR14Q catalog.
A global continuum including a power-law and an iron emission template pseudo continuum \citep{Boroson1992,Vestergaard2001,Salviander2007} is fitted with tens of separated line-free spectral windows, and the monochromatic luminosity (at $\lambda = $ 5100\r{A} ($L_{5100}$), 3000\r{A} ($L_{3000}$), and 1350\r{A} ($L_{1350}$) for quasars at various redshifts) is then derived.
The host galaxy contamination is not considered as most of our sources are high redshift luminous quasars (see S11).

Following S11, we fit various emission lines separately.
For \Hb~(of objects with $z\le 0.89$), we use a power-law continuum with iron template to fit within the wavelength windows of [4435, 4700]\r{A} and [5100, 5535]\r{A}.
The emission lines are fitted within [4700, 5100]\r{A} with 8 Gaussians, 
three for the broad components of \Hb\ with Full Width at Half Maximum (FWHM) $>$ 1200 \kms, one for the narrow component with FWHM $<$ 1200 \kms, and the rest four for the narrow \OIII\ $\lambda \lambda$4959,5007 doublet (one core and one wing component for each line). 
The FWHMs and the velocity offsets of \Hb\ narrow line and the \OIII\ core component are tied up.
The same restriction is also applied on the wing component of \OIII.
For \MgII\ ($0.35 \le z$ $\le$ 2.25), we fit the continuum spectra utilizing the same continuum model as above, but within the spectral windows of [2200, 2700]\r{A} and [2900, 3090]\r{A}, and the line over [2700, 2900]\r{A} with three Gaussians for the broad component and one for narrow.
For \CIV~($z \ge 1.5$), only a power-law continuum is used over windows of [1445, 1465]\r{A} and [1700, 1705]\r{A}, and only 3 Gaussians for broad component over the spectral range of [1465, 1700]\r{A}.
We note that, for simplicity and to be easy to reproduce, the numbers of broad Gaussians we used for the lines are fixed rather than variable in S11. 
To ensure the reliability of the fitting, we only adopt the results of lines with a median spectral S/N (signal-to-noise ratio) around the line-fitting region $>$ 3, which is roughly corresponded to a $\pm 20\%$ fitting bias of FWHM and EW for high-EW objects \citep{Shen2011}. 
We finally stress that hereafter, unless otherwise stated, the \Hb\ or \MgII\ line refers to the broad component we derive from the spectral fitting. 

Using the best-fit broad Gaussian models, we measure for each line the EW and FWHM with {\tt PyQSOFit}, and the line asymmetry with Pearson's skewness coefficient: ${\rm skewness}=3(\lambda_{\rm mean} - \lambda_{\rm median})/\sigma_{\lambda}$ \citep{VandenBerk2001}.
The $\lambda_{\rm median}$ is where the wavelength bisects the area of the emission line model while the $\lambda_{\rm mean}$ is defined as 
\begin{equation}
   \lambda_{\rm mean}=\frac{\int_{-\infty}^{+\infty}{\lambda f(\lambda) d\lambda}}{\int_{-\infty}^{+\infty}{f(\lambda) d\lambda}}.
\end{equation}
We stress that the skewness under such definition can reveal the shape of the emission line model only, regardless of the systematical shift.

We estimate the black hole (BH) mass of quasars based on single-epoch spectrum assuming virialized BLR (S11). 
With the continuum luminosity as a proxy for the BLR radius and the FWHM of broad line as a proxy for virial velocity, the virial BH mass can be given following the expression as 
\begin{equation}\label{equation:virialBHmass}
   \log\left({\frac{M_{\rm BH}}{M_{\odot}}}\right)=a+b\log\left({\frac{\lambda L_{\lambda}}{10^{44}~\ergs}}\right)+2\log\left({\frac{\rm FWHM}{\kms}}\right),
\end{equation}
where the $L_{\lambda}=L_{5100}$ for \Hb, $L_{\lambda}=L_{3000}$ for \MgII\, and $L_{\lambda}=L_{1350}$ for \CIV.
We adopt the calibration parameters in S11 (cf. their Equations 5, 8, and 6, respectively):
\begin{eqnarray} 
   \begin{tabular}{cc}\label{EQ:Hb}
      (a, b)=(0.910, 0.50), &\Hb
   \end{tabular}\\
   \begin{tabular}{cc}
      (a, b)=(0.740, 0.62), &\MgII
   \end{tabular}\\
   \begin{tabular}{cc}
      (a, b)=(0.660, 0.53), &\CIV
   \end{tabular}
\end{eqnarray} 
We also adopt the same bolometric corrections (BCs) as S11 to estimate the bolometric luminosity ($L_{\rm bol}$) where ${\rm BC}_{5100}=9.26$, ${\rm BC}_{3000}=5.15$, and ${\rm BC}_{1350}=3.81$, respectively.

\begin{figure*}[tb!]
    \includegraphics[width=1\textwidth]{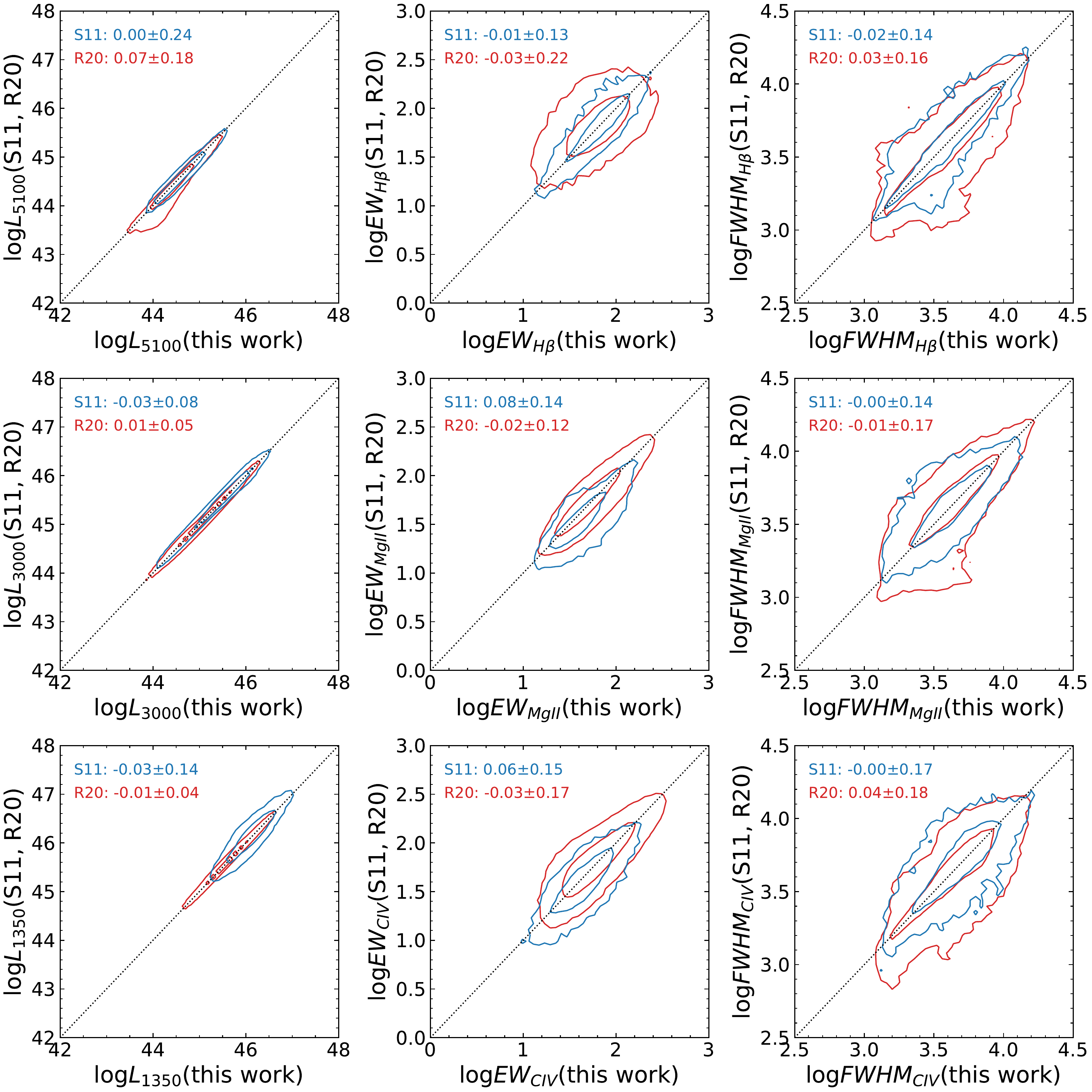}
    \caption{Comparisons of the continuum luminosity, line EW and FWHM between S11, R20, and this work.
The mean and standard deviations of the difference are nominated in the upper left corner of each panel. In each panel, the inner and outer contours are the $1\sigma$ and $2\sigma$ density contours, respectively. \label{fig:compcatalog}}
\end{figure*}

We note that a recent work has compiled spectral properties as well as SMBH mass and bolometric luminosity measurements for the DR14Q quasars \citep[][hereafter R20]{Rakshit2019} using the same {\tt PyQSOFit} code and similar parameters.
In R20, the continuum components are fitted to the whole spectrum with various templates, while in this work, we prefer to fit the continuum with nearby line-free spectral windows.
Comparing to the global fit in R20, the local fit perform better in fitting continuum flux with simple models within a limited region, so that it could provide more precise emission line spectra.
Besides, as we will state below, in this work we are more interested in spectra in the most extreme state for EVQs, 1,363 of which are not the primary thus not included in R20.
In addition, in this work, detailed spectral measurements (such as line skewness, bisectional line center and the properties of composed spectra) are required.
Therefore we choose our independent spectral fitting results in this work for the following analyses. 

Nevertheless, we present in Fig. \ref{fig:compcatalog} the comparison of our measurements of luminosity, line EW and FWHM with those of R20 and as well as S11, 
showing rather negligible deviations and small scatter between the measurements. 

We will not go deeper discussing the details of the weak deviation and smaller scatter between the measurements, however, adopting the measurements from R20 does not alter the results presented in this work.

\section{The EVQ samples and control samples} \label{sec:sample}

\subsection{EVQs Selection} \label{subsec:EVQ_sel}

\begin{figure}[tb!]
  \includegraphics[width=.48\textwidth]{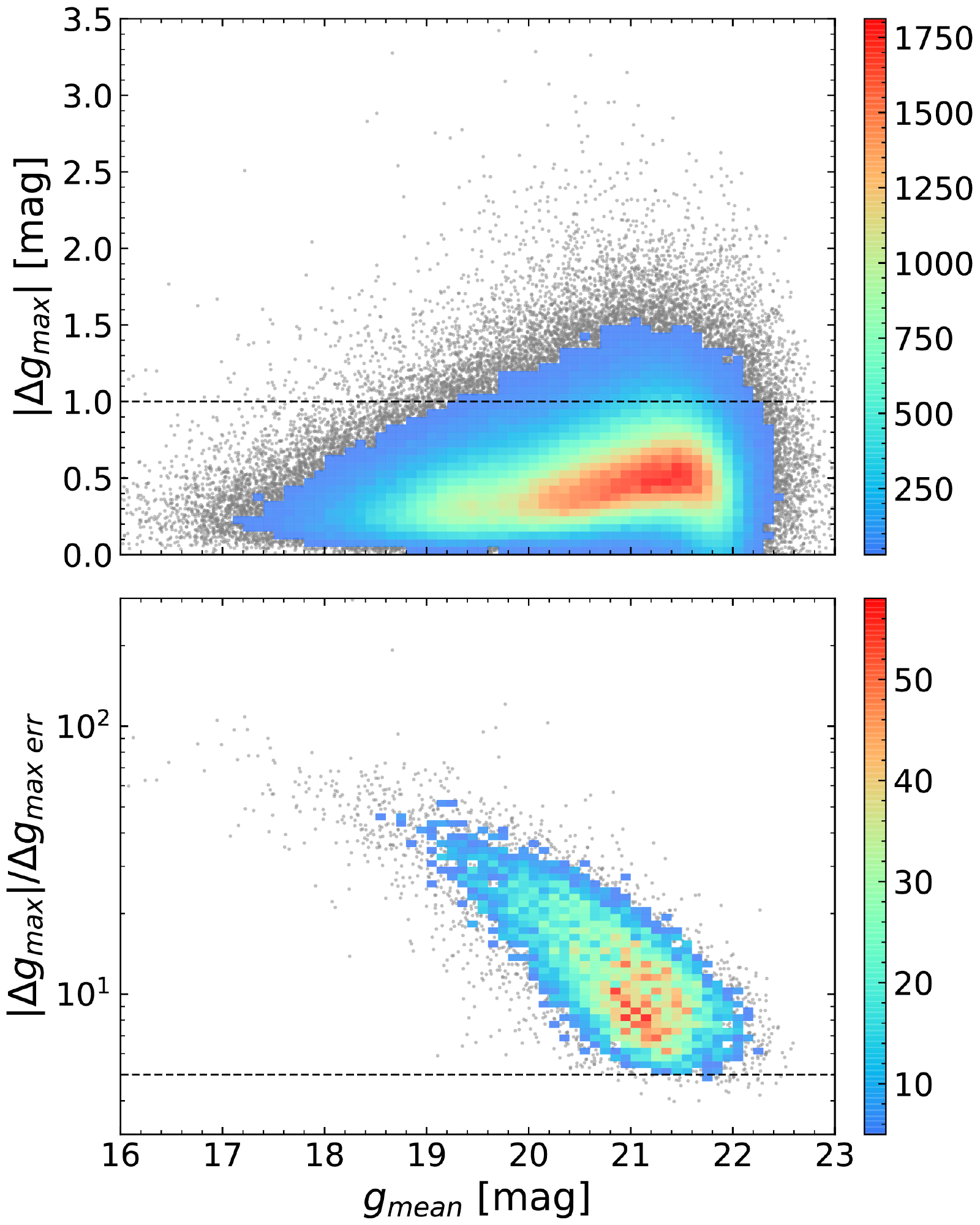}
  \caption{Upper: $|\Delta g_{max}|$ versus $g_{mean}$ of all quasars, with the black dashed line marks the criteria we set for EVQs in $g$-band. Lower: the significance of $|\Delta g_{max}|$ of the EVQs we selected in \S\ref{subsec:EVQ_sel}. The dashed line is plotted at 5$\sigma$.
  \label{fig:vardist}}
\end{figure}

\begin{figure}[tb!]
   \includegraphics[width=.48\textwidth]{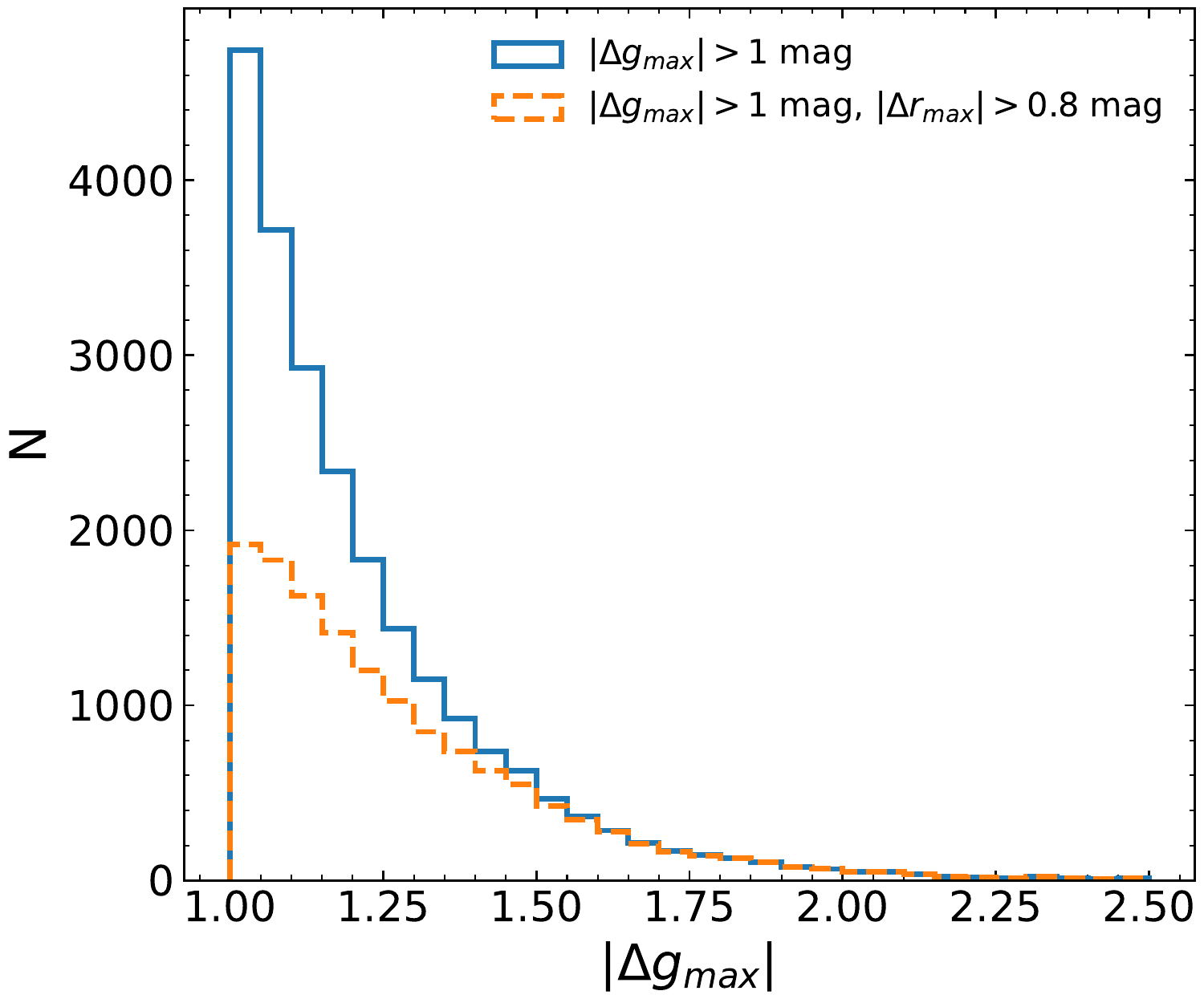}
   \caption{The distribution of $|\Delta g_{max}|$ of EVQs selected by the criterion in this work and a single $|\Delta g_{max}|>1$ mag.
   \label{fig:evqcriterion}}
\end{figure}

Using the clean and paired $g$- and $r$-band light curve we derived in \S\ref{subsec:rejectbad}, we consider a source as an EVQ if any of its two photometry pairs satisfied $|\Delta g_{max}|>1$ mag and $|\Delta r_{max}|>0.8$ mag quasi-simultaneously.
By this criterion, 14,012 EVQs are selected, a catalog of them will be released in a future paper in this series.
We plot $|\Delta g_{max}|$ versus $g_{mean}$ of the sample in the upper panel of Fig \ref{fig:vardist}. Though most EVQs are faint sources, most of their $|\Delta g_{max}|$ are statistically significant (with S/N $>$ 5, lower panel of Fig \ref{fig:vardist}), as we have dropped photometric data points with magnitude error $>$ 0.2 mag (see \S\ref{sec:data}).

We note that a simultaneous $|\Delta r_{max}|>0.8$ mag adopted in this work is a strong and conservative request, rejecting which would yield 22,740 candidates instead\footnote{This number is still considerably smaller than $\sim$ 56000 aforementioned if applying $|\Delta g_{max}|>1$ mag on the $g$-band light curves before the ``cleaning" process described in \S\ref{subsec:rejectbad}. This is because the ``cleaning" process has excluded potentially problematic epochs which could yield spuriously large $\Delta g$ (see Fig. \ref{fig:detectiondiff}), and also a portion of $g$-band data points without paired (obtained within 30 days) $r$-band PS1 exposures (see Fig.\ref{fig:epochinterval}).}. 
Practically, EVQs with $|\Delta g_{max}|>1$ mag and $|\Delta r_{max}|>0.8$ mag simultaneously have more extreme variability than those selected with a single $|\Delta g_{max}|>1$ mag criterion. This could be clearly seen in Fig. \ref{fig:evqcriterion}. 

\begin{figure}[tb!]
   \includegraphics[width=.48\textwidth]{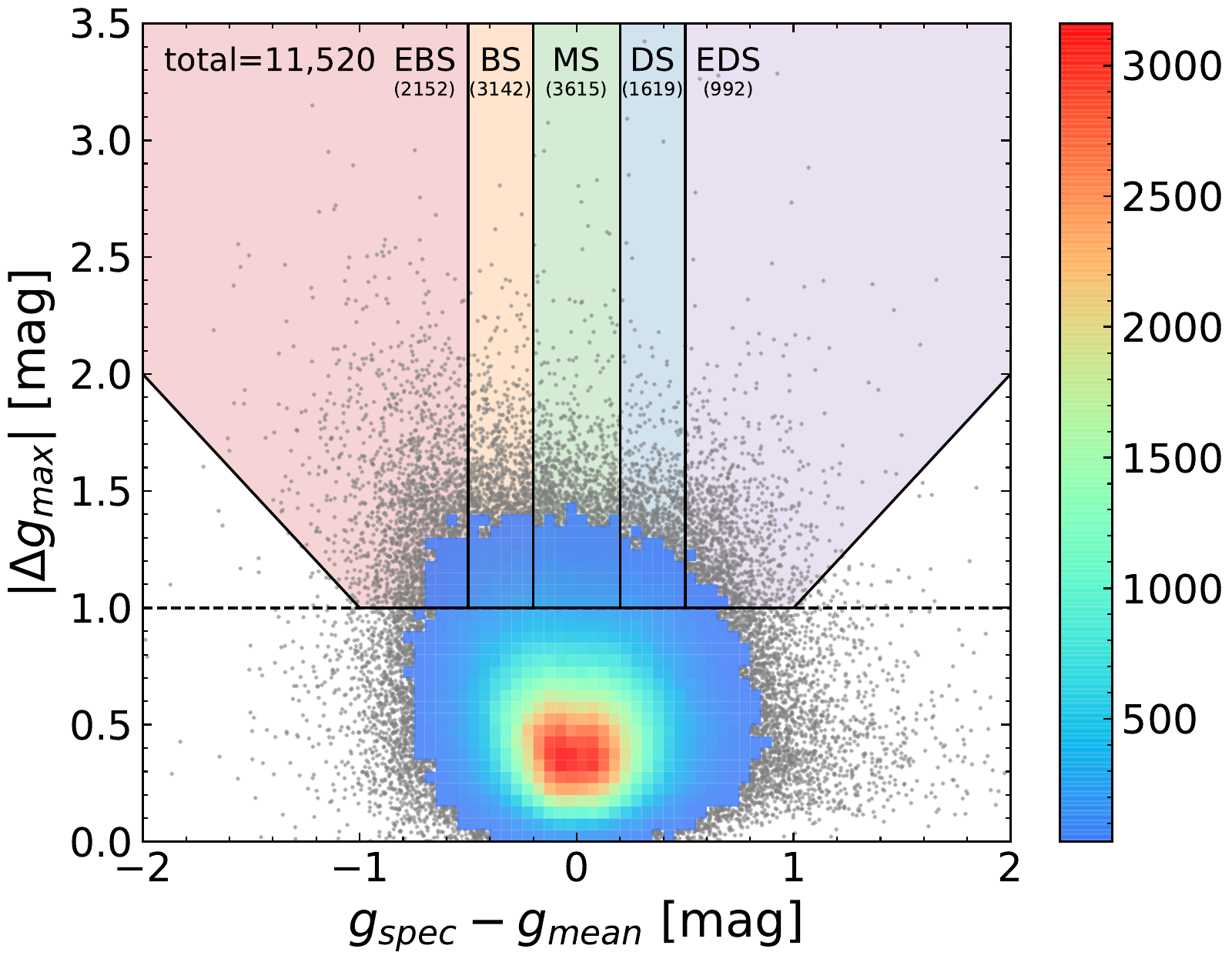}
   \caption{Criteria used to sort the spectral states of EVQs. $|\Delta g_{max}|$ is the largest change in $g$-band photometric magnitude for each source and the $g_{\rm spec}-g_{\rm mean}$ represents the deviation of the $g$-band synthetic spectral magnitude from the $g$-band mean photometric magnitude. See \S\ref{subsec:EVQ_sel} for details.
   \label{fig:evqsamp}}
\end{figure}

Before we determine the state of the spectra of EVQs, we first exclude 3,991 low quality spectra ($g$-band synthetic magnitude error $<0.2\ {\rm mag}$), leaving 2,447 EVQs with no available spectra.
To parameterize the states of the spectra, we calculate the magnitude difference between their synthetic photometry $g_{\rm spec}$ derived from SDSS spectra (the AB magnitude evaluated from the \textit{spectroFlux} given by SDSS which was derived through convolving the spectrum with the corresponding filter) and the mean photometric magnitude ($g_{\rm mean}$, see \S\ref{subsec:photometry}).
To derive the stacked spectra and explore possible variation of the spectral feature in different states, re-binning the sample is necessary.
According to the commonly used EVQs criterion ($|\Delta g_{max}|>1$), we class the spectra with $|g_{\rm spec}-g_{\rm mean}|>0.5$ into extreme states (symmetrically into extremely bright state or extremely dim state). We further divide the rest 8,376 spectra into three classes (see below for details), so that, the whole sample is divided into five classes, 
which enable us to explore the gradual variation of spectral features from extremely dim to extremely bright states. 
On the whole, the criterion can be expressed as follow:

\begin{enumerate}
  \item $-\Delta g_{max} <g_{\rm spec}-g_{\rm mean}<-0.5$ : Extremely Bright State (EBS);
  \item $-0.5 \le g_{\rm spec}-g_{\rm mean}<-0.2$ : Bright State (BS);
  \item $-0.2 \le g_{\rm spec}-g_{\rm mean}<0.2$ : Median State (MS);
  \item $0.2 \le g_{\rm spec}-g_{\rm mean}<0.5$ : Dim State (DS);
  \item $0.5 \le g_{\rm spec}-g_{\rm mean}<\Delta g_{max}$ : Extremely Dim State (EDS).
\end{enumerate}
where the $\Delta g_{max}$ is the greatest magnitude change in each $g$-band photometric light curve.
A sketch of the above criteria is shown in Fig. \ref{fig:evqsamp}.
For those 2,341 EVQs with multiple SDSS spectra, we only keep the most extreme spectrum the $g$-band synthetic magnitude of which departs most from the mean photometric magnitude from SDSS and PS1. We defer the study of spectral variability in these individual EVQs with multiple spectra to a future work in this series.
We note that dividing the parent sample into more classes would reduce the number of quasars in each class, and adopting different number of classes or using boundaries different from what we adopt would not alter the results in this work.
Furthermore, 
there could be other strategies to re-bin the EVQs into various states. For instance, one may choose to re-bin according to ($g_{\rm spec}-g_{\rm mean}$)/$\Delta g_{max}$ (instead of $g_{\rm spec}-g_{\rm mean}$, see \S\ref{sec:rebin}), i.e., to normalize the magnitude deviation $g_{\rm spec}-g_{\rm mean}$ by the maximum variability amplitude even seen. However, $\Delta g_{max}$ only represents the maximum variability amplitude of a quasar seen by SDSS and PS1 photometric survey (i.e., would be significantly affected by the sampling), but not necessary an intrinsic physical property of the quasar. Meanwhile, this alternative strategy would not alter the results in this work either.

There are a few EVQs with too bright (13) or too dim (32) synthetic magnitude ($|g_{\rm spec}-g_{\rm mean}| > \Delta g_{max}$).
We note that the SDSS spectra taken by recent BOSS campaigns have a smaller fiber diameter (2\arcsec\ vs 3\arcsec) than the former SDSS campaigns. 
The smaller fiber might increase the possibility of the fiber-drop that results in biased spectra with significantly low flux density which has been reported in literature \citep[e.g., ][]{Shen2015, Sun2015, Guo2020}. 
More information about fiber-drop can be found in \cite{Dawson2013}.
We suspect the majority of the too dim spectra of 32 EVQs (4 from SDSS-I and II and 28 from SDSS-III and IV) were due to fiber-drop, and exclude them from this study. The too bright spectra of 13 EVQs may be physically real signals or spurious (such as due to natural or artificial solar system objects accidentally passing through the line of sights during spectroscopic observations). Presently, we also exclude them from this paper, and defer studies on these individual sources to a future work. 

In total, 11,520 EVQs remain, including 2,152 in EBS, 3,142 in BS, 3,615 in MS, 1,619 in DS, and 992 in EDS.
In the following analyses, in order to avoid confusion caused by plotting too many subsamples in a single plot, when necessary we also merge the EBS/BS subsamples into ABS (all bright states) and the EDS/DS into ADS (all dim states).
We note that there are considerably fewer sources in DS (EDS) compared with BS (EBS). 
That is because a clear portion of sources that should belong to DS (EDS) are excluded for their poor synthetic magnitude.
Nevertheless, even if we keep those sources with too faint spectra, the dim samples are still considerably smaller. 
This could be a selection bias\footnote{See \cite{Shen2021} for a different manifestation of this kind of bias caused by AGN variability.} that if a quasar was in a dim state, it may not have been spectroscopically identified, or even may not have been selected for spectroscopic observation, and would be missed by the quasar catalogs.
However such biases would not affect the following analyses in this work, as the control samples (see \S\ref{subsec:control} below) are built to have matched redshift and spectra-derived monochromatic luminosity (thus matched brightness in the spectra) with our EVQ samples.  
\begin{figure*}[tb!]
  \includegraphics[width=1\textwidth]{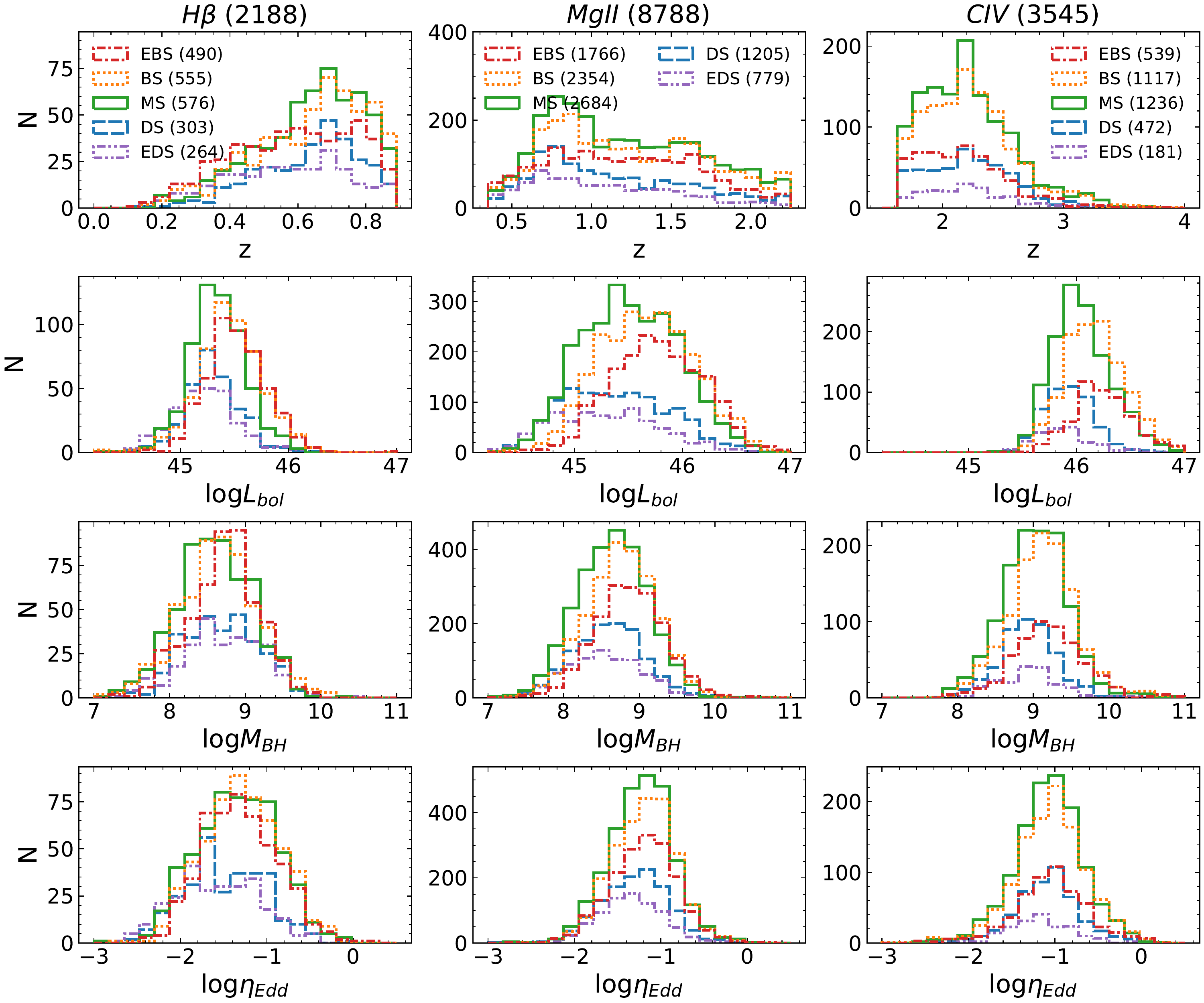}
  \caption{The distribution of redshift $z$, $L_{\rm bol}$, $M_{\rm BH}$ and $\eta_{\rm Edd}$ of our EVQ samples. Numbers in brackets in the plot indicate the size of each sample or sub-sample. \label{fig:sample}}
\end{figure*}

\subsection{Control Samples} \label{subsec:control}

We then build control samples of subtly-variable quasars with essentially the same physical parameters including redshift, bolometric luminosity and SMBH mass with our EVQs, for the comparison of the spectra properties between EVQs and the control samples.
The matching in redshift, bolometric luminosity and SMBH mass is critical as the spectral properties of quasars could significantly evolve (intrinsically or due to intricate observational effects) with such parameters.

The subtly-variable quasars samples are selected out of quasars with
$|\Delta g_{max}|<0.4$ and also $|g_{\rm spec}-g_{\rm mean}|<0.4$ (around 140,000 quasars satisfy such criteria).
We build the first set of control samples using the $L_{\rm bol}$ and $M_{\rm BH}$ of EVQs measured directly from their single-epoch spectra (\S\ref{subsec:specfit}),
by selecting the most alike subtly-variable quasar for each EVQ in the space of $L_{\rm bol}$, $M_{\rm BH}$ and redshift (hereafter direct control sample, or DCS).
Naturally, each DCS control sample has the same size and matched redshift, $L_{\rm bol}$, $M_{\rm BH}$, broad line FWHM, and also the Eddington ratio ($\eta_{\rm Edd}$) as those of its corresponding EVQ sample.

However, EVQs are experiencing extreme variations, thus the bolometric luminosity derived from the single-epoch spectrum must have been biased by such variations, particularly for those quasars with spectra captured in their extreme states. To better represent the long-term average brightness of a quasar we apply a correction to the aforementioned single epoch bolometric luminosity ($L_{\rm bol}$, and also the monochromatic luminosity $L_{\lambda}$) for each quasar: 
\begin{equation}
   \log{L_{\rm bol,``cor"}}=\log{L_{\rm bol}}+\frac{g_{\rm spec}-g_{\rm mean}}{2.5}
\end{equation}
We then build the 2nd set of control samples (luminosity-``corrected" control samples, hereafter LCS) with matched redshift, $L_{\rm bol,``cor"}$ and single-epoch SMBH mass. Note during the correction we simply assume the variability amplitude of $L_{\rm bol}$ is the same as that of $g$ band luminosity. Alternatively, we may simply request matching in average $g$ band luminosity (effectively in $g_{\rm mean}$ since redshift is also matched), which however would not alter the results in this work.

Besides, the single-epoch virial black hole mass estimates may also biased by variability (the breathing of broad line region, i.e., the change of line width with luminosity in individual AGNs). It was found while H$\beta$ line display normal breathing expected from the virial relation \citep{Gibson2008, Denney2009,Park2012, Barth2015a, Runco2016}, \MgII\ shows much weaker breathing \citep[e.g.,][]{Shen2013, Yang2020a, Guo2020a, Homan2020}, and \CIV\ exhibits even anti-breathing \citep[e.g.,][]{Richards2002a, Wilhite2006, Shen2008a, Sun2018d, Wang2020}. Simply assuming no breathing of \MgII\ and \CIV\ (i.e., the line width does not vary with luminosity)\footnote{The broad line breathing in EVQs could be explored using individual EVQs with multiple spectra. We would defer this to a future work in this series.}, we could further apply a correction to the single-epoch SMBH mass through replacing $L_{\lambda}$ with $L_{\lambda,``cor"}$ during the calculation of mass in Equation \ref{equation:virialBHmass}:
\begin{equation}\label{equation:viralmass2}
   \log\left({\frac{M_{\rm BH, ``cor"}}{M_{\odot}}}\right)=a+b\log\left({\frac{\lambda L_{\rm \lambda, ``cor"}}{10^{44}~\ergs}}\right)+2\log\left({\frac{\rm FWHM}{\kms}}\right).
\end{equation}
The 3rd set of control samples (for \MgII\ and \CIV\ samples only) are selected to have matched redshift, $L_{\rm bol,``cor"}$ and $M_{\rm BH, ``cor"}$ (Mass-``corrected" control samples, MCS hereafter).
Further note that it has been shown $M_{\rm BH}$ estimated using high-ionization emission lines like \CIV\ \citep[e.g., ][]{Sulentic2007,Shen2012,Runnoe2013,Coatman2016,Coatman2017} should be used with cautious as \CIV\ may not be virialized. In this case, the control samples could be interpreted more precisely as matched in \CIV\ line FWHM (but not necessarily in SMBH mass). 

Note neither the LCS nor MCS samples are perfect control samples of our EVQs, also because the broad line breathing, the bolometric correction factors and the calibration parameters for the virial mass might be different for EVQs and normal quasars, or vary from source to source.
However, further meticulous fix is out of our scope and seems unnecessary as we will show later our results in this work are insensitive to the choice of the three control samples we built.

\begin{figure}[tb!]
    \includegraphics[width=.48\textwidth]{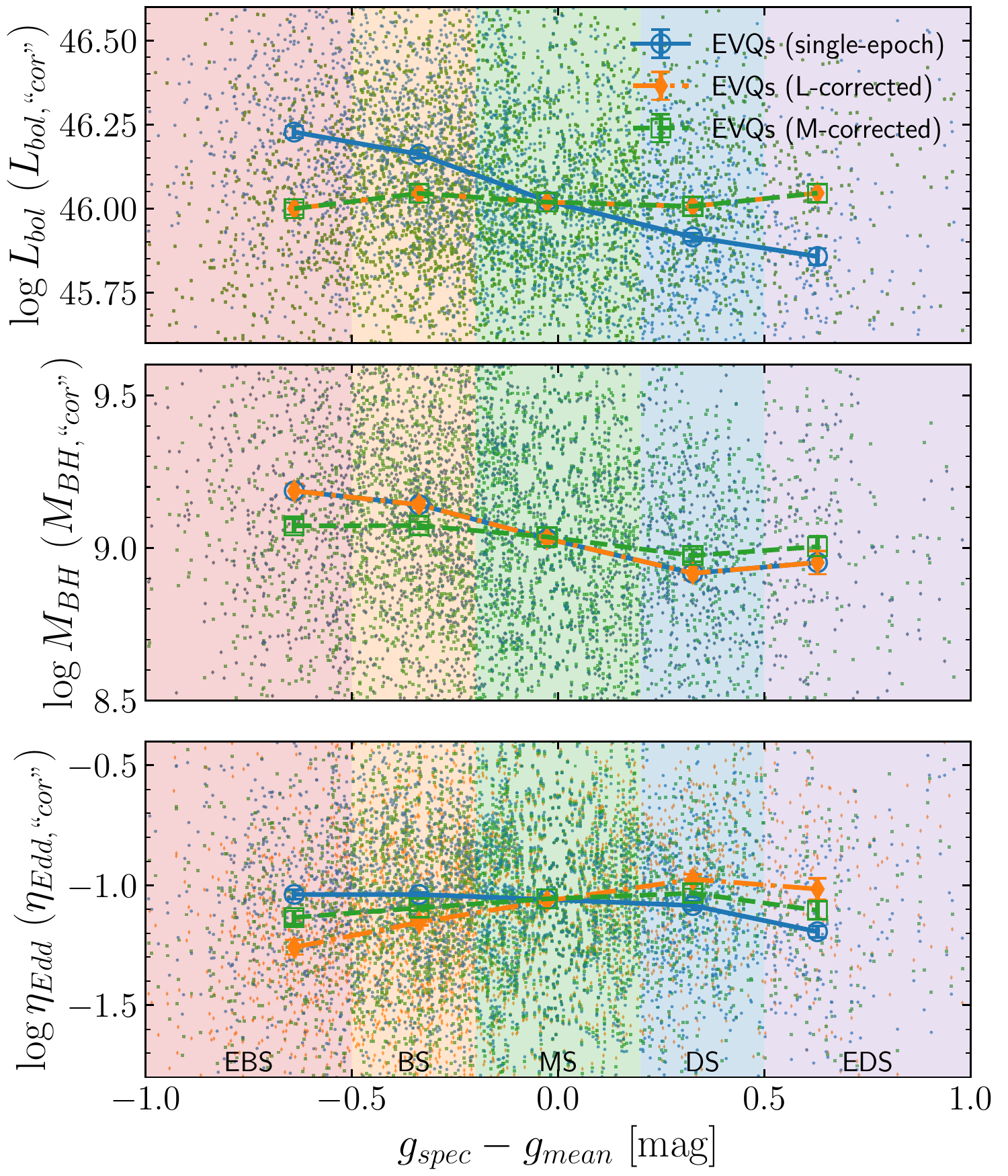}
    \caption{The single-epoch $L_{\rm bol}$ ($L_{\rm bol,``cor"}$), $M_{\rm BH}$ ($M_{\rm BH,``cor"}$) and $\eta_{\rm Edd}$ ($\eta_{\rm Edd,``cor"}$) versus $g_{\rm spec}-g_{\rm mean}$ of our EVQs and control samples (using the \CIV\ sample for example). To demonstrate the difference between EVQs and their control samples, we plot the median values of EVQs in different spectral states (see \S\ref{subsec:EVQ_sel} for the re-binning criteria) and of their corresponding control samples. The 1$\sigma$ errors in the medians are derived from bootstrapping the corresponding samples.
    \label{fig:correction}}
\end{figure}

\section{Results} \label{sec:results}

Due to the limited SDSS spectral coverage and the fact we utilize monochromatic luminosity at different wavelengths and different broad lines to derive the bolometric luminosity and the virial black hole mass for different redshift ranges, in this work, we treat the \Hb\ sample (z $<$ 0.89), \MgII\ sample (0.35 $\leq$ z $<$ 2.25) and \CIV\ sample (1.5 $\leq$ z $<$ 4) separately. 
In Fig. \ref{fig:sample} we plot the distribution of the redshift, single-epoch $L_{\rm bol}$, $M_{\rm BH}$ and $\eta_{\rm Edd}$ of our EVQ samples. 
The distributions of $z$, $L_{\rm bol}$ or $L_{\rm bol,``cor"}$, $M_{\rm BH}$ or $M_{\rm BH,``cor"}$ of the control samples, which are not plotted here, are statistically indistinguishable from their corresponding EVQ samples according to the K-S test. 

We present Fig. \ref{fig:correction} to illustrate the effects of luminosity-``correction" and mass-``correction" (using the CIV sample for example).
In Fig. \ref{fig:correction} we find that EVQs in bright states tend to have higher median single-epoch $L_{\rm bol}$ and subsequently higher $M_{\rm BH}$, compared with those in dim states.
The luminosity-``correction”, which was proposed to derive the long-term averaged luminosity of a quasar, yields a similar median $L_{\rm bol,``cor"}$ for all EVQ sub-samples. 
Similarly, mass-``correction” gives mass based on $L_{\rm bol,``cor"}$ (instead of $L_{\rm bol}$), thus yields median $M_{\rm BH,``cor"}$ less sensitive (compared to $M_{\rm BH}$) to EVQ state. 

\subsection{The Composite Spectra} \label{compspec}

\begin{figure*}[tb!]
   \includegraphics[width=1\textwidth]{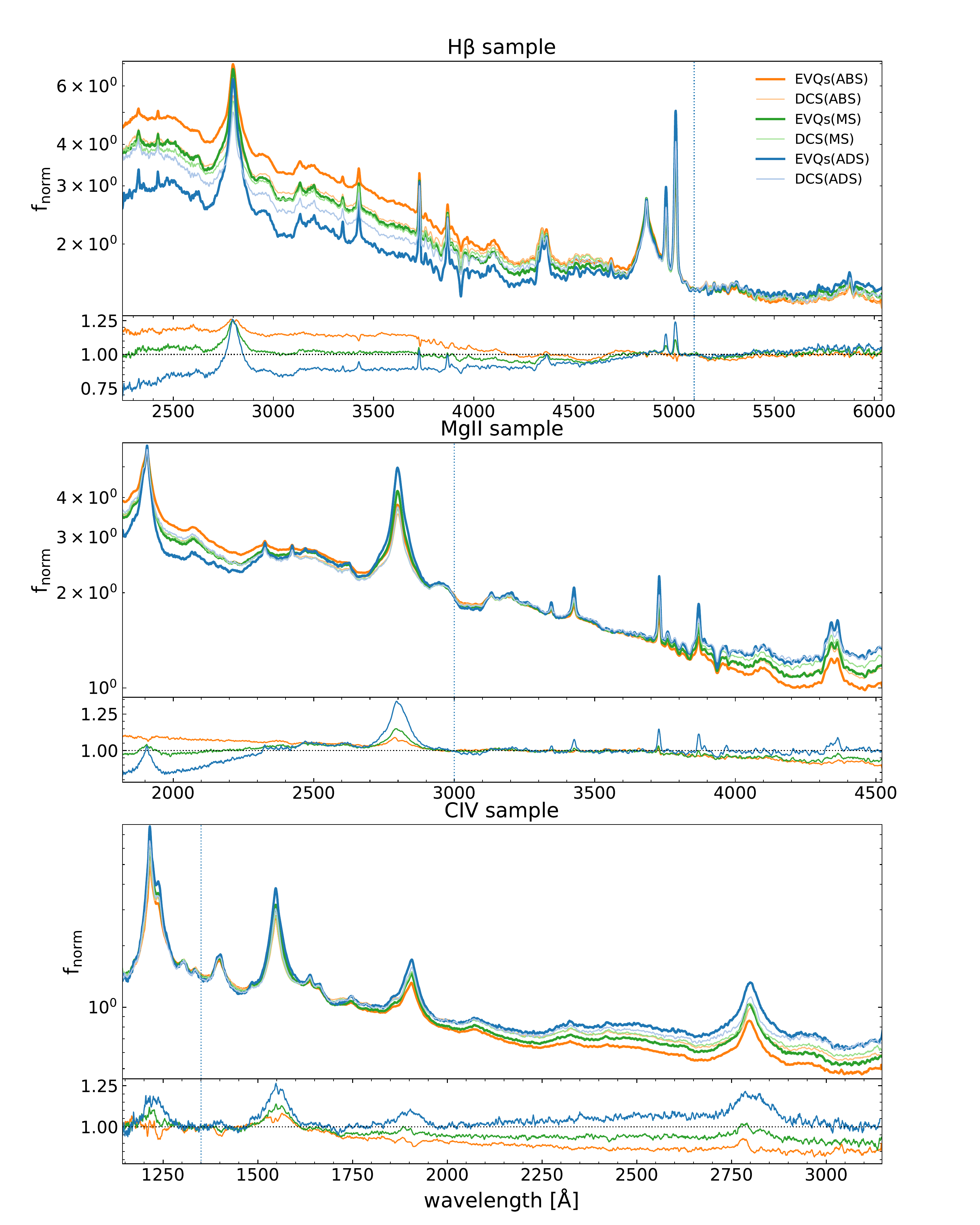}
   \caption{The stacked (geometric mean) spectra and the spectral ratio of EVQs and their corresponding direct control samples. EVQs(ABS): in all bright states, including EBS and BS; EVQs(MS): in median states; EVQs(ADS): in all dim states, including EDS and DS.
   The vertical dashed lines mark the wavelengths we adopted to normalize the spectra. The composite spectra are cut at blue and red ends where $<$ 5\% of sources in the sample contribute to the stacking.
   \label{fig:compareconti}}
\end{figure*}

To display the overall spectra of EVQs and the control samples, we construct composite spectra for each sample aforementioned. 
Before stacking, each spectrum is corrected for Galactic extinction and shifted to the rest frame.
For each sample, we normalize the spectra from various quasars at the wavelength of which we derive the monochromatic luminosity (see \S\ref{subsec:specfit}, also marked in Fig. \ref{fig:compareconti}) and derive a geometric mean spectrum.
To avoid confusion in the plots, we merge EVQs in Bright State and Extremely Bright State into EVQs(ABS), and those in Dim State and Extremely Dim State into EVQs(ADS). Their corresponding control samples are equally treated. The composite spectra of EVQs and control samples and plotted in Fig. \ref{fig:compareconti}. 

Compared with the control samples, EVQs in brighter states exhibit clearly bluer spectra, and reversely redder spectra in dim states.
The trend is consistent with the so-called ``bluer-when-brighter" pattern widely seen in AGNs and quasars (see \S\ref{subsec:bwb} for discussion).

From Fig. \ref{fig:compareconti} we also see clear line residuals in the spectra ratios of EVQs and the control samples, particularly for EVQ(ADS) and EVQ(Median), showing EVQs tend to have stronger emission lines compared with their control samples. We present detailed comparison of the line EW in \S\ref{subsec:EW} and line profile in \S\ref{subsec:profile}. 

Note in Fig. \ref{fig:compareconti} we only plot the DCS control samples (to avoid confusion). Replacing DCS samples with LCS/MCS samples will not alter the results.
A minor note is that the DCS control samples for ABS, MS, and ADS exhibit somehow slightly different spectral slopes between themselves, but only significant in the low redshift bin (\Hb\ sample). This is likely because these direct control samples for EVQs have lower bolometric luminosity in dim states compared with those in bright states,
that the relatively stronger host contamination yields a redder spectrum\footnote{The difference is much weaker or disappears if we instead plot LCS/MCS samples which show indistinguishable luminosity distributions.}.
The effect of host contamination is much weaker at shorter rest-frame wavelength thus the difference almost disappears in higher redshift bins, which is also consistent with previous studies which found that near UV spectral slopes of quasars show little dependence on luminosity \citep[e.g.][]{Telfer2002, Bonning2007}.

\subsection{Line Equivalent Widths}\label{subsec:EW}
\begin{figure}[tb!]
    \includegraphics[width=.48\textwidth]{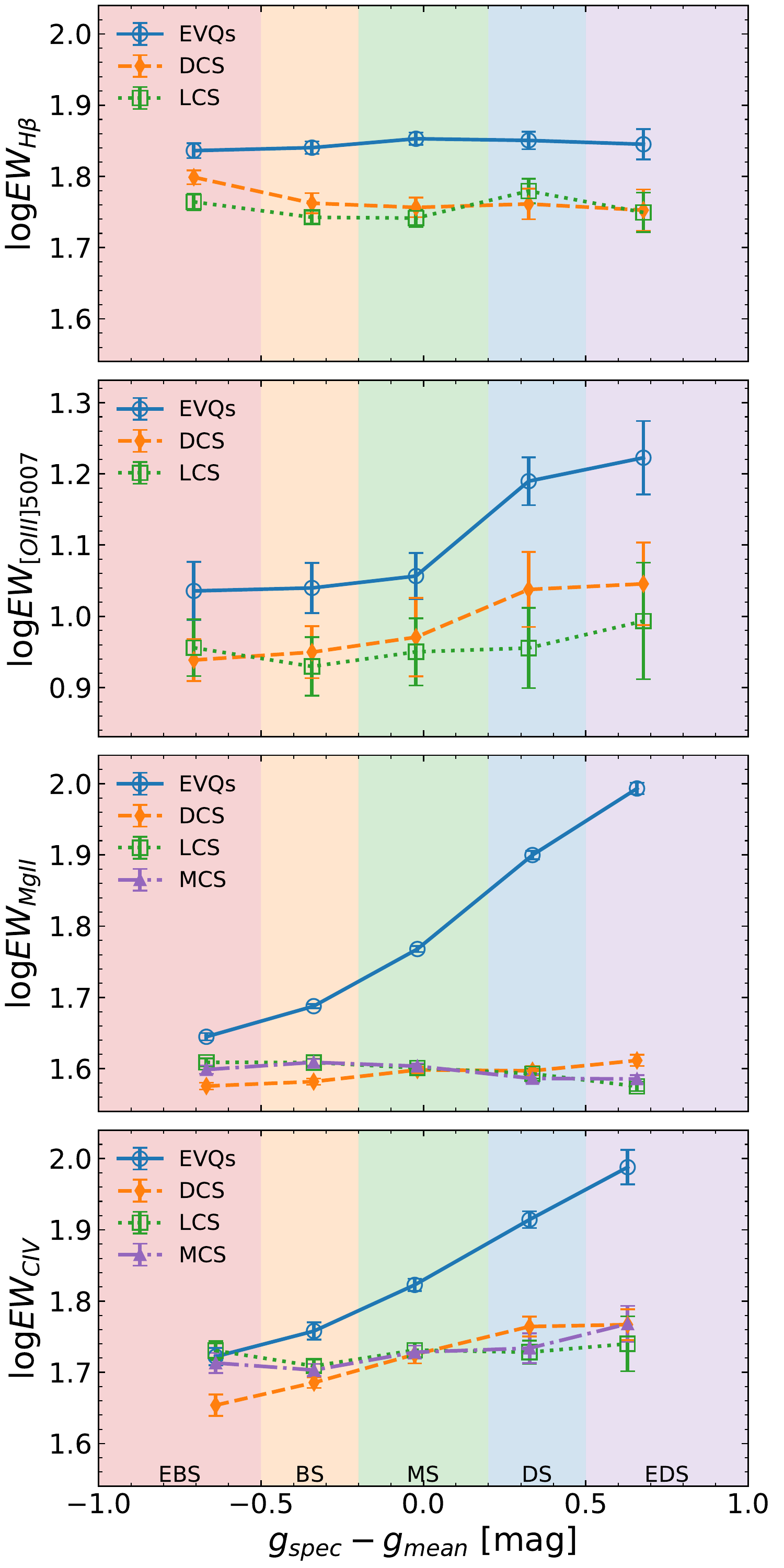}
    \caption{Emission line EWs measured from the stacked spectra of EVQs in different states, and of the corresponding control samples. 
    The error bars are derived from bootstrapping the corresponding sample used to derive the stacked spectra.
    \label{fig:ewstack}}
\end{figure}
\begin{figure*}[tb!]
    \includegraphics[width=1\textwidth]{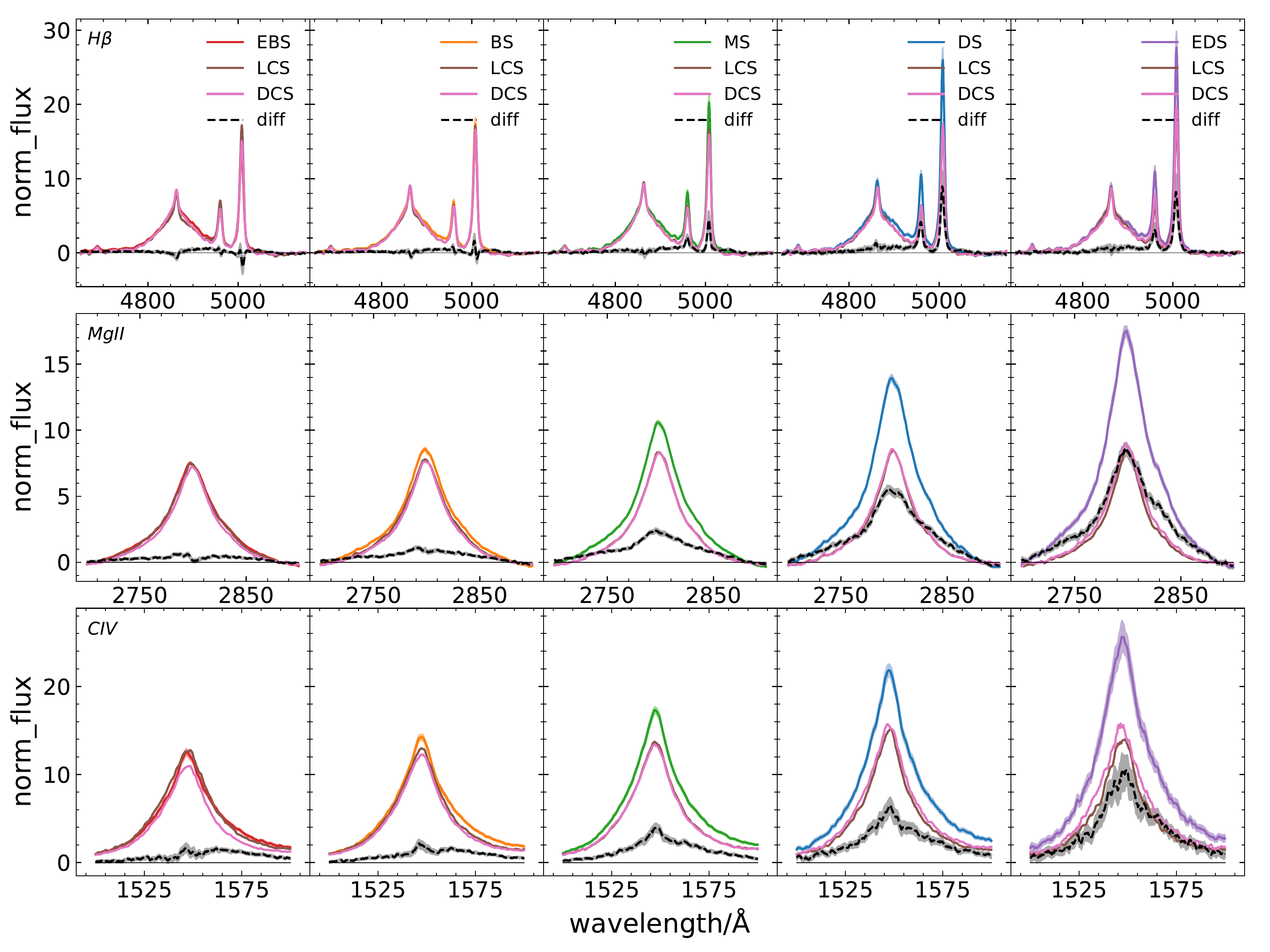}
    \caption{The median stacked emission line spectra, normalized by the continuum flux density at the corresponding line center (\Hb, \MgII\ and \CIV\ respectively, marked in the left upper corner of each row), and with the best-fit continuum models subtracted. 
    The black dashed lines are the difference spectra between EVQs and DCS.
    The black horizontal lines at zero are over-plotted for reference. 
    We also plot with shaded spectra the 1$\sigma$ errors (derived through bootstrapping the corresponding samples) of the stacked spectra of EVQs and of the difference spectra between EVQs and DCS.
    \label{fig:line_conti_norm}}
\end{figure*}

Following the procedures we adopt to fit the individual spectra, we also fit the composite spectra to derive the line parameters.
To illustrate the difference of the line EW between EVQs and their control samples,
we plot in Fig. \ref{fig:ewstack}
the best-fit line EWs (of \OIII, \CIV, broad \MgII\ and \Hb) derived from the stacked spectra of EVQs in various states (and of their control samples).

Clearly, using the control samples as references (the main results we present below are indeed insensitive to the choice of control samples), we find that EVQs in their extremely dim and dim states tend to have larger line EWs and contrarily smaller EWs in their extremely bright and bright states.
This could be primarily be attributed to the so-called intrinsic Baldwin effect (iBeff), i.e., emission line EW in individual AGNs often decreases when AGNs brighten \citep[e.g.][]{Pogge1992,Kinney1990,Homan2020}.

Puzzlingly, the broad \Hb\ line behaves differently. 
Though the iBeff has been clearly detected in individual AGNs \citep{Goad2004, Rakic2017}, the stacked spectra of our EVQs exhibit no iBeff of broad \Hb. 

We further find that, from the overall trend, EVQs tend to have larger emission line EWs compared with the control samples though the extent varies from dim to bright states due to the iBeff.
By comparing the EW in MS which is free from the iBeff, we find that the EW of broad \MgII\ in EVQs is higher by $\sim$ 47\% compared with the control sample.
For broad \Hb, \OIII\ and \CIV, the numbers are $\sim$ 27\%, $\sim$ 25\% and $\sim$ 25\% respectively.
Spectra ratio plots in Fig. \ref{fig:compareconti} also reveal clear line residuals around \OIII, \MgII\ and \CIV\ line, further demonstrating that EVQs have systematically larger line EWs compared with control samples. 
A minor note is that, in Fig. \ref{fig:compareconti} 
we could barely see line residuals around \Hb.
The larger best-fit EW of \Hb\ in EVQs shown in Fig. \ref{fig:ewstack} might be due to the excess of the very broad component of \Hb\ in EVQs (see \S\ref{subsec:profile}).

\subsection{Line Profiles} \label{subsec:profile}

\begin{figure*}[tb!]
    \includegraphics[width=1\textwidth]{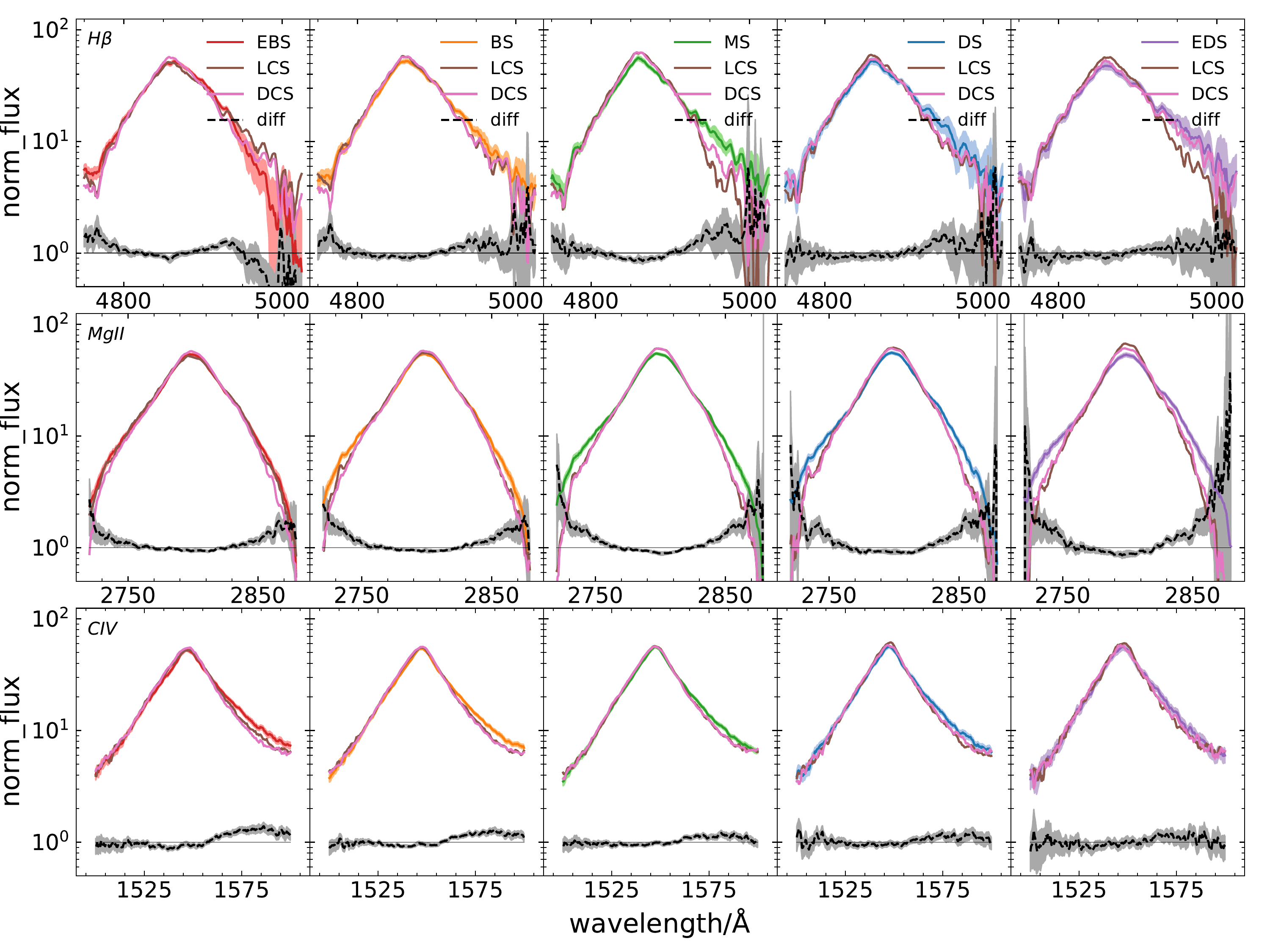}
    \caption{Similar to Fig. \ref{fig:line_conti_norm}, but the median stacked emission line spectra are plotted in logarithm space, with both the best-fit continuum and narrow line models subtracted,  and normalized by the line fluxes (of broad \Hb, \MgII, and \CIV\ respectively). The black dashed lines are the ratio spectra of EVQs and DCS. The black horizontal lines at unity are over-plotted for references. \label{fig:line_flux_norm}}
\end{figure*}
We plot in Fig. \ref{fig:line_conti_norm} the stacked emission line spectra of our EVQs in various states, in comparison with the control samples. Following \cite{VandenBerk2001}, the stacked emission line spectra were obtained through normalizing each spectrum by the continuum flux density at the corresponding line center, median stacking the spectra, fitting the median spectra and subtracting the continuum models. 
From Fig. \ref{fig:line_conti_norm} we could barely see differences in the line profiles between EVQs and the control samples. 
This is actually expected as the control samples were built to have matched luminosity and SMBH mass with EVQs (thus matched broad line FWHM as the SMBH mass is a virial product of continuum luminosity and line FWHM).
Note the situation is slightly different for LCS. LCS samples were built to have matched long-term averaged luminosity and single-epoch spectroscopy-based SMBH mass (compared with EVQs), thus the matching in line FWHM is not guaranteed. In fact, the LCS samples for EVQs in bright (dim) states tend to have slightly larger (smaller) broad line FWHMs. However as we will show below, such an effect is weak and negligible.
For MCS, since we use the long-term averaged luminosity to derived SMBH mass, matching in broad line FWHM is also guaranteed.
We again stress that the results we provide below are insensitive to the choice of the control samples.

To further explore potential subtle differences in the line profiles between EVQs and the control samples, we plot in Fig. \ref{fig:line_flux_norm} the broad line spectra normalized by accumulated line fluxes. 
The stacked line profile of broad \Hb\ and \MgII\ seem symmetric, and the \CIV\ line exhibit clear redward skewness.
 Through plotting the spectra in logarithm space in Fig. \ref{fig:line_flux_norm}, we find EVQs tend to have stronger broad line wings compared with the control samples. The excesses in such broad-base components are generally seen in all three broad lines we concern (\Hb, \MgII, and \CIV) and all states of EVQs. They seem symmetric in \MgII\ (with excesses seen in both the blue and red wings), but redward asymmetric in \CIV\ (only seen in the red wing) and probably also in the dim states of \Hb. 
 
\begin{figure}[tb!]
   \includegraphics[width=.48\textwidth]{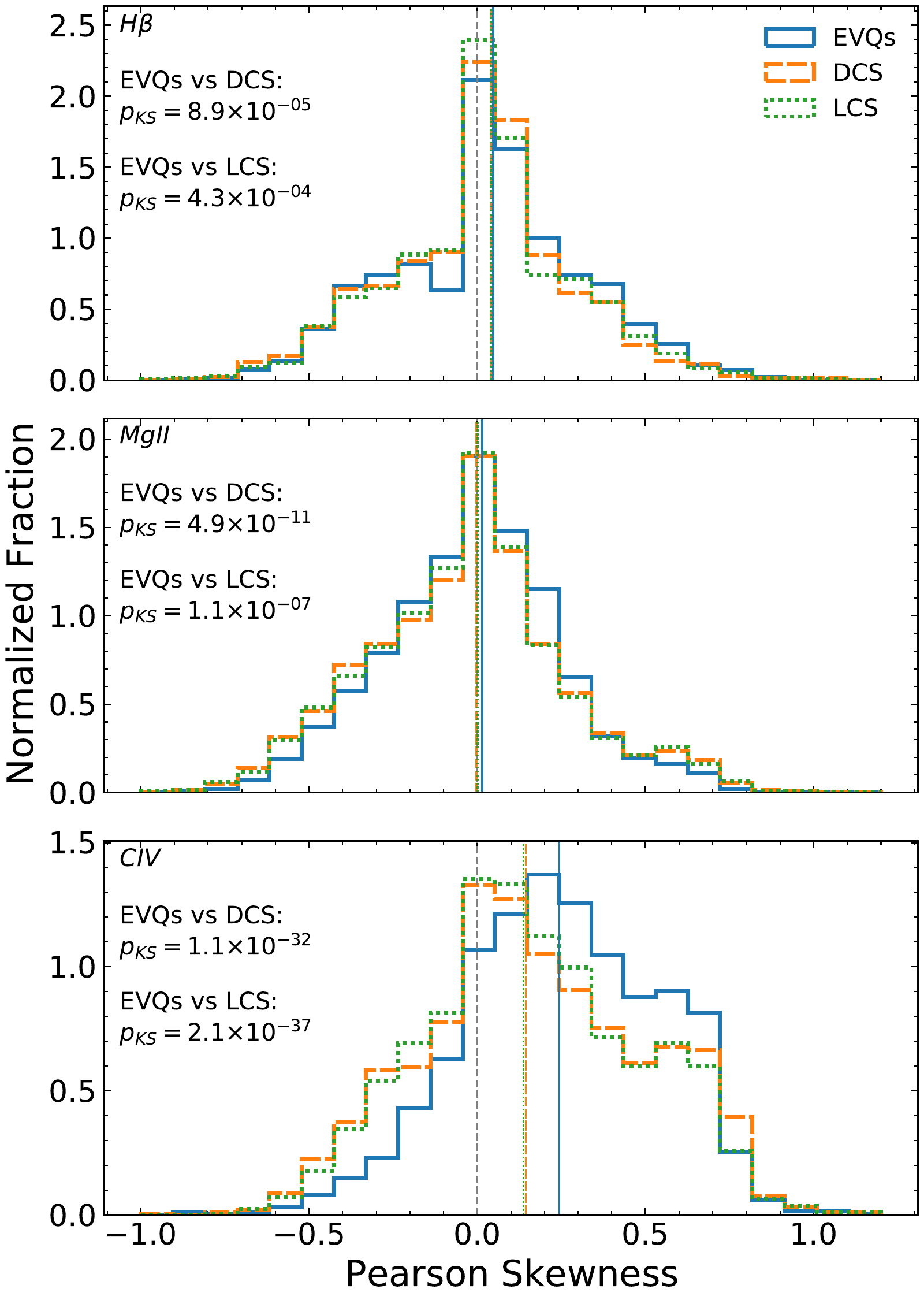}
   \caption{Distribution of the Pearson skewness of broad \Hb, \MgII\ and \CIV\ emission lines in EVQs and control samples. The colored dashed lines indicate the median value of each sample and a line at zero for reference. The p-value of the K-S test between EVQs and the control samples are given.
   \label{fig:civskewness}}
\end{figure}

We also plot in Fig. \ref{fig:civskewness} the distribution of Pearson skewness measured from individual sources for our EVQs and control samples. 
The patterns illustrated in Fig. \ref{fig:civskewness} are consistent with what we have revealed from the stacked line profiles (Fig. \ref{fig:line_flux_norm}). The median skewness values of broad \Hb\ and \MgII\ are much closer to zero than that of \CIV\, showing \Hb\ and \MgII\ are mainly symmetric while \CIV\ exhibits redward asymmetry. 
Because of the clear excess of the redshifted broad-base component,
the \CIV\ line of EVQs is dramatically more redward skewed compared with the control samples, and the K-S test gives $p=1.1 \times 10^{-32}$ between EVQs and DCS and $p=2.1 \times 10^{-37}$ between EVQs and LCS;
The K-S test (see Fig. \ref{fig:civskewness}) also reveals statistical difference between the skewness parameter distributions between EVQs and their control samples for \Hb\ and \MgII\, suggesting \Hb\ and \MgII\ are also slightly more redward in EVQs, though their median skewness values (of EVQs and control samples) are very close.

To explore the contribution of the excess quantitatively,
the broad lines are further fitted with core and wing components.
We use two Gaussian with FWHM $<$ 6000 \kms\ to represent the core component and one Gaussian with FWHM $>$ 6000 \kms\ for the very broad component (wing). The stacked spectra of EVQs in each state and of their corresponding control samples are fitted together with the line widths and center linked and normalization free to vary.
As a result, we find the very broad component in EVQs accounts for higher fraction of the total line flux compared with their control samples. 
The fraction contributed by the very broad component in EVQs (and in their control samples) to total line flux is 
$\sim$ 55.0\% (39.5\%) for \Hb, $\sim$ 47.6\% (36.1\%) for \MgII\ and $\sim$ 55.2\% (47.6\%) for \CIV, respectively.

\section{Discussion} \label{sec:discussion}

In this work we build a large sample of EVQs and divide them into sub-samples according to their brightness states during spectroscopic observations. We carefully build control samples with matched redshift, luminosity, and SMBH mass.
The comparison between EVQs and such control samples enables us to probe the nature and consequences of the extreme variability precluding potential effects of these parameters.
The control samples (with matched redshift, spectroscopic monochromatic luminosity, and line FWHM) also enable the comparison free from intricate observational biases, e.g., spectroscopic identification of quasars and spectral fitting may rely on spectral quality and line FWHM.

\subsection{The ``bluer-when-brighter" pattern}\label{subsec:bwb}

Through comparing the composite spectra of EVQs at various states, we find that
EVQs follow the general ``bluer-when-brighter" pattern widely seen in quasars and AGNs \citep[e.g.][]{Cutri1985,Wamsteker1990,Clavel1991,Giveon1999,Webb2000,Trevese2001,VandenBerk2004,Wilhite2005,Meusinger2010,Sakata2011,Schmidt2012,Bian2012,Zuo2012,Sun2014,Ruan2014,Cai2016,Guo2016,Cai2019SCPMA,Guo2020}.
The color variation is actually also visible in Fig. \ref{fig:detectiondiff}, where we could see that EVQs (which contribute blue dots in the green zones) follow the same $\Delta r$ vs $\Delta g$ trend of normal quasars. The best orthogonal regression slope ($\Delta r$ = 0.878 $\Delta g$) is less than unity, also demonstrating a ``bluer-when-brighter" pattern (similarly, see Fig. 1 in \citealt{Rumbaugh2018}, but from a much smaller sample).

The similar ``bluer-when-brighter" trend seen in EVQs and normal quasars suggests a common physical origin of the variation. Historically, the ``bluer-when-brighter" behaviour had been attributed to host galaxy contamination \citep[e.g.][]{Choloniewski1981, Hawkins2003}, or changes in global accretion rate \citep[e.g.][]{Pereyra2006}.
Such models have been refuted by recent observations \citep[e.g.][]{Schmidt2012,Sun2014,Zhu2016}.
It is now more widely accepted that UV/optical variations are due to magnetic turbulence in the accretion disc \citep[e.g.][]{Kelly2009}, and the flux and color variations could be modelled with thermal fluctuations \citep{Dexter2011, Ruan2014, Cai2016, Cai2018a, Cai2020}. The same mechanism may also be responsible for the extreme variability seen in EVQs.

\subsection{The Intrinsic Baldwin effect}\label{subsec:iBeff}

\begin{figure}[tb!]
   \includegraphics[width=.48\textwidth]{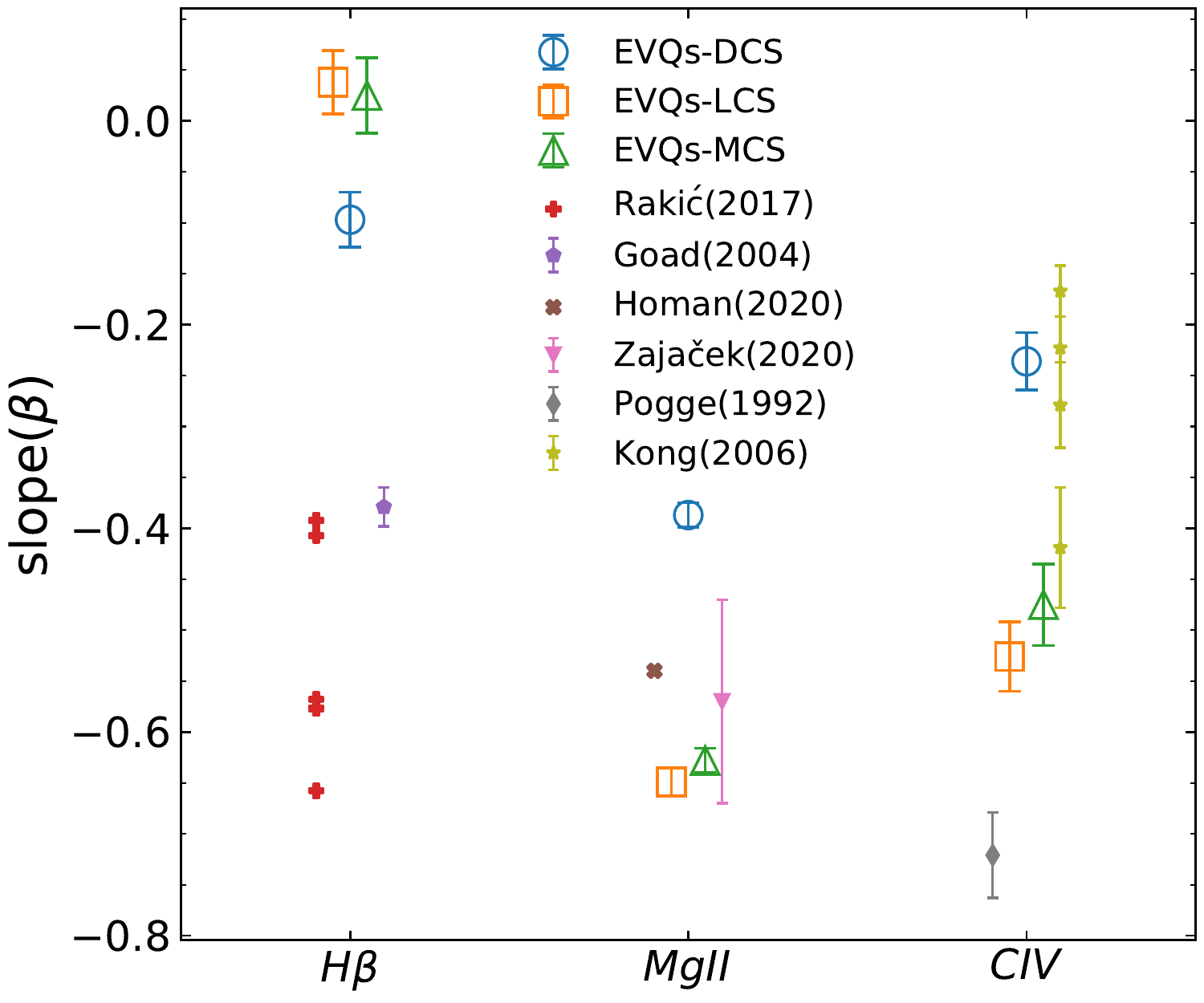}
   \caption{The mean slope of ``iBeff" of EVQs (derived throgh compare EVQs in various states) of broad \Hb, \MgII, and \CIV. As comparison, we also plot the iBeff slopes measured from individual AGNs \citep{Pogge1992, Goad2004, Kong2006, Rakic2017, Zajacek2020} and an AGN sample \citep{Homan2020} with multiple spectra. 
   \label{fig:ibeff}}
\end{figure}

\begin{deluxetable*}{cccccc}
    \tablenum{1}
    \tablecaption{Linear regression and Spearman's rank correlation results of the iBeff of EVQs \label{tab:ibeff}}
    \tablewidth{0pt}
    \tablehead{
        \colhead{} & \colhead{sample} & \colhead{$\beta$} & \colhead{$\alpha$} & \colhead{Spearman's rank $\rho$} & \colhead{$p$}
    }
    \decimalcolnumbers
    \startdata
        \multirow{3}{*}{\Hb} & EVQs-DCS & $-0.097\pm0.030$ & $0.054\pm0.006$ & -0.063 & 0.003\\
        & EVQs-LCS & $0.038\pm0.038$ & $0.060\pm0.006$ & 0.018 & 0.401\\
        & EVQs-MCS & $0.025\pm0.036$ & $0.053\pm0.007$ & 0.009 & 0.670\\
        \hline
        \multirow{3}{*}{\MgII} & EVQs-DCS & $-0.387\pm0.014$ & $0.143\pm0.003$ & -0.286 & $\ll10^{-10}$\\
        & EVQs-LCS & $-0.649\pm0.015$ & $0.141\pm0.003$ & -0.443 & $\ll10^{-10}$\\
        & EVQs-MCS & $-0.628\pm0.014$ & $0.139\pm0.003$ & -0.439 & $\ll10^{-10}$\\
        \hline
        \multirow{3}{*}{\CIV} & EVQs-DCS & $-0.236\pm0.040$ & $0.107\pm0.005$ & -0.109 & $\ll10^{-10}$\\
        & EVQs-LCS & $-0.526\pm0.030$ & $0.105\pm0.005$ & -0.237 & $\ll10^{-10}$\\
        & EVQs-MCS & $-0.475\pm0.033$ & $0.107\pm0.005$ & -0.231 & $\ll10^{-10}$\\
    \enddata
    \tablecomments{The iBeff slope of EVQs presented in Fig. \ref{fig:ibeff}. (3) and (4) are the best-fit slope $\beta$ and the intercept $\alpha$ in Equation \ref{equation:ibeff}. (5) and (6) are the Spearman's rank correlation coefficient $\rho$ and the corresponding confidence level p-value.}
\end{deluxetable*}

The iBeff can be expressed by a simple formula $\rm{EW}\propto L_{cont}^\beta$, where $L_{cont}$ is the continuum luminosity \citep[e.g.,][]{Kinney1990, Pogge1992}.
In this work, using the control samples of EVQs as reference, we estimate the mean iBeff slope ($\beta$) of EVQs by linear fitting the data with the following relation:
\begin{equation}\label{equation:ibeff}
    \log{\rm{\frac{EW_{EVQs}(i)}{EW_{ctrl}(i)}}}=-\beta~\frac{g_{spec}(i)-g_{mean}(i)}{2.5}+\alpha,
\end{equation}
where $EW_{EVQs}(i)/EW_{ctrl}(i)$ is the ratio of the line EW of an EVQ to that of the corresponding normal quasar in the control sample, and ${g_{spec}(i)-g_{mean}(i)}$ is the deviation of its synthetic spectroscopy magnitude from its mean photometric magnitude. 
The derived iBeff slopes are presented in Table \ref{tab:ibeff} and Fig. \ref{fig:ibeff}. In Table \ref{tab:ibeff} we also present the Spearman’s rank correlation coefficients between the two quantities on each side of Equation \ref{equation:ibeff}, which yield correlation patterns consistent with the results from the linear regression.

We note that the iBeff slopes of EVQs derived using the DCS as reference are considerably different from those using LCS or MCS as reference. This is particularly prominent for \MgII\ and \CIV\ that using DCS as reference yields much flatter iBeff slopes. 
This is because EVQs in dim and extreme dim states have systematically lower luminosity compared with EVQs in bright and extreme bright states (see Fig. \ref{fig:correction}), thus their DCS control samples would suffer the ensemble Baldwin effect (see Fig. \ref{fig:ewstack}).
Clearly here using LCS or MCS as reference is less biased to probe the iBeff of EVQs here. 

We see significant iBeff of \MgII\ and \CIV\ in our EVQs, and the iBeff slopes are comparable to those reported in literature for individual AGNs with multiple spectra.
Statistically, the slope of \MgII\ iBeff is slightly steeper than that of \CIV, which is contrary to the trend seen in eBeff that the eBeff is stronger for lines with higher ionization energy \citep{Dietrich2002},
which might indicate that the physical origins between the iBeff and eBeff are not connected \citep{Rakic2017}. 
One possibility is that \MgII\ line is produced at larger distance and does not respond to continuum variation as fast as \CIV\ thus stronger iBeff is expected.

However, it is rather puzzling that the broad \Hb\ line of EVQs does not show clear iBeff.
This is not only contrary to \MgII\ and \CIV\ of EVQs, but also to the clear iBeff of \Hb\ commonly detected in individual AGNs.
For instance, a study on a few (six) long-term monitored AGNs revealed the iBeff of \Hb\ in all subsets of type 1 AGNs (i.e. Seyfert 1, narrow line Seyfert 1 or high-luminosity quasars) with $\beta\gtrsim -0.4$ \citep{Rakic2017}.
But note the reported iBeff slopes in literature vary from source to source, or even from year to year for the same source \citep[][]{Rakic2017,Goad2004}.
We notice that the host galaxy contamination could reduce the continuum variability amplitude thus alter the iBeff slope (see \S3.6 in S11 for relevant discussion of the role of host contamination on the ensemble Baldwin effect). However, even if after we restrict to EVQs at $\log{L_{bol}}>45.8$ (the most luminous $\sim$10\% EVQs in the \Hb\ sample), we still get rather weak iBeff (with slopes well above -0.12).

\Hb\ line is previously known to be exceptional in the ensemble Baldwin effect. 
The eBeff always occurs in high-ionization emission lines whose correlation will be steeper when ionization energy goes higher \citep{Dietrich2002}.
However, the Balmer lines, like \Ha\ and \Hb, exhibit no correlation \citep[e.g.,][]{Kovacevic2010,Popovic2011} or even a weak positive correlation \citep{Croom2002a,Greene2005a} between the EW and luminosity of continuum, although \Lya\ line does exhibit eBeff \citep[e.g.][]{Dietrich2002}.
Meanwhile, a recent work of \cite{Kang2021} found that while more variable quasars have clearly stronger (with larger EW) \MgII, \OIII\ and \CIV\ lines (after correcting the effects of bolometric luminosity, black hole mass and redshift), broad \Hb\ shows rather weaker correlation between EW and UV/optical variability amplitude.
This trend is also visible in Fig. \ref{fig:ewstack} and \ref{fig:line_conti_norm} in this work, that while EVQs have systematically stronger \MgII\ and \CIV\ line compared with the control samples, the difference is less prominent for broad \Hb.
It is yet unclear why \Hb\ behaves exceptionally.
\cite{Dietrich2002} proposed that the lack of eBeff of Balmer lines might be due to the intricate and unclear radiation processes of Balmer lines \citep{Netzer2020}.

We finally note that it would be intriguing to examine the iBeff of individual EVQs with multiple spectra. We would defer this to a future paper of this series.

\subsection{EVQs have stronger emission lines} \label{subsec:largerew}

In Fig. \ref{fig:ewstack} we see EVQs have systematically larger line EWs compared with their control samples. The difference is highly prominent for EVQs in extremely dim state because of the iBeff, remains statistically significant in medium state, and is even visible in the extremely bright state for \MgII. 
A similar conclusion, free from our sample division, can also be found in Table \ref{tab:ibeff} where we fit the correlation between the $\log{EW_{EVQs}/EW_{ctrl}}$ and the ${g_{spec}-g_{mean}}$ of our EVQs. Comparing with all three control samples, the intercepts are well above zero in the three emission line samples, further support our findings in stacked spectra.

\cite{Rumbaugh2018} has reported that EVQs seem to have systematically larger EW in UV emission lines (compared with normal quasars with matched redshift and luminosity), and attributed such phenomenon to the overall lower Eddington ratio of EVQs.
However, in this work, we find EVQs have stronger emission lines, even compared with a group of control sample with matched luminosity and SMBH mass thus matched $\eta_{\rm Edd}$.
The results are unaltered if we rectify the measurements of bolometric luminosity and the virial black hole mass which might be biased by the extreme variability (see the comparison with LCS and MCS in Fig. \ref{fig:ewstack}).

The discovery that EVQs have systematically stronger emission lines is in good agreement with a recent study of \cite{Kang2021}, who found the UV/optical variation amplitude of quasars in SDSS Stripe positively correlate with emission line EWs\footnote{\cite{Kang2021} has shown that the correlation between broad \Hb\ line EW and variability amplitude is however much weaker. This is also consistent with the pattern shown in our Fig. \ref{fig:compareconti} and \ref{fig:ewstack}. }, after controlling the effects of redshift, luminosity, and SMBH mass. One possible explanation for such correlation is that stronger disc fluctuations could lead to harder quasars SED \citep{Cai2018a}, thus provide relatively more ionizing photons.  Alternatively, stronger disc turbulence may be able to launch BLR clouds with larger sky coverage. 

Notably, \cite{Ross2020} reported several \CIV\ changing look quasars, which show intrinsic Baldwin effect of \CIV\ line, and have larger \CIV\ line EWs comparing with sources with matched SMBH mass (see Fig. 3 of \citealt{Ross2020}). 
Therefore, we can see that normal quasars, EVQs, and CLQs exhibit similar intrinsic Baldwin effect, and they follow the same trend that more variable quasars tend to have systematically larger line EWs. These facts suggest common physical processes behind these various populations. 

\subsection{The excess of very broad line component} \label{subsec:redexcess}

Comparing the broad line profiles of EVQs with their control samples,
we find EVQs exhibit subtle excess of the very broad line component (VBC, see Fig. \ref{fig:line_flux_norm}). The excess is similarly visible in all states of EVQs. 

Statistical studies on line profiles had suggested the broad line region of quasars consists of two components: a very broad line region (VBLR) closer to the central SMBH and an intermediate line region (ILR) at larger distance \citep{Wills1993,Brotherton1994,Sulentic2000}.
The existence of multiple BLR components is further supported by variation studies which revealed distinct variation patterns of the two line components \citep[e.g.][]{Sulentic2000ApJ,Hu2020,Guo2020}.
Since the VBLR lies closer to the SMBH, the VBC could easily reverberate (respond to the variation of the central ionizing continuum) in short time. Contrarily, the ILR may appear non-reverberating or reverberating at much longer time. The observed ``anti-breathing'' of \CIV\ line (i.e., the line broadens when luminosity increases) could also be attributed to the combination the two components \citep[e.g., ][]{Denney2012,Wang2020}.

The likely physical origin of the VBLR is optically thin gas located near the black hole \citep{Popovic1995, Corbin1997, Corbin1997a}.
The Keplerian velocity of that gas could lead to a very broad line width and the SMBH gravity could yield systematical line redshift.
The excess of the VBC we discover in EVQs suggests that the strong disc turbulence associated with the extreme variability could launch more gas into VBLR from the accretion disk. 

The excess of the VBC is clearly redward skewed in \CIV, but not in \MgII\ and \Hb. This is likely because the dominant ILR \CIV\ flux comes from the accretion disk wind which is significantly outflowing thus yielding systematically blue-shifted emission (see also \S\ref{subsec:blueshift}), while the VBLR is not outflowing. 
Furthermore, the gravitational redshift of the VBLR region could be more prominent for \CIV\ which could locate at smaller radii compared to \MgII\ and \Hb.

\subsection{Radio Loudness} \label{subsec:radio}

\begin{figure}[tb!]
   \includegraphics[width=.48\textwidth]{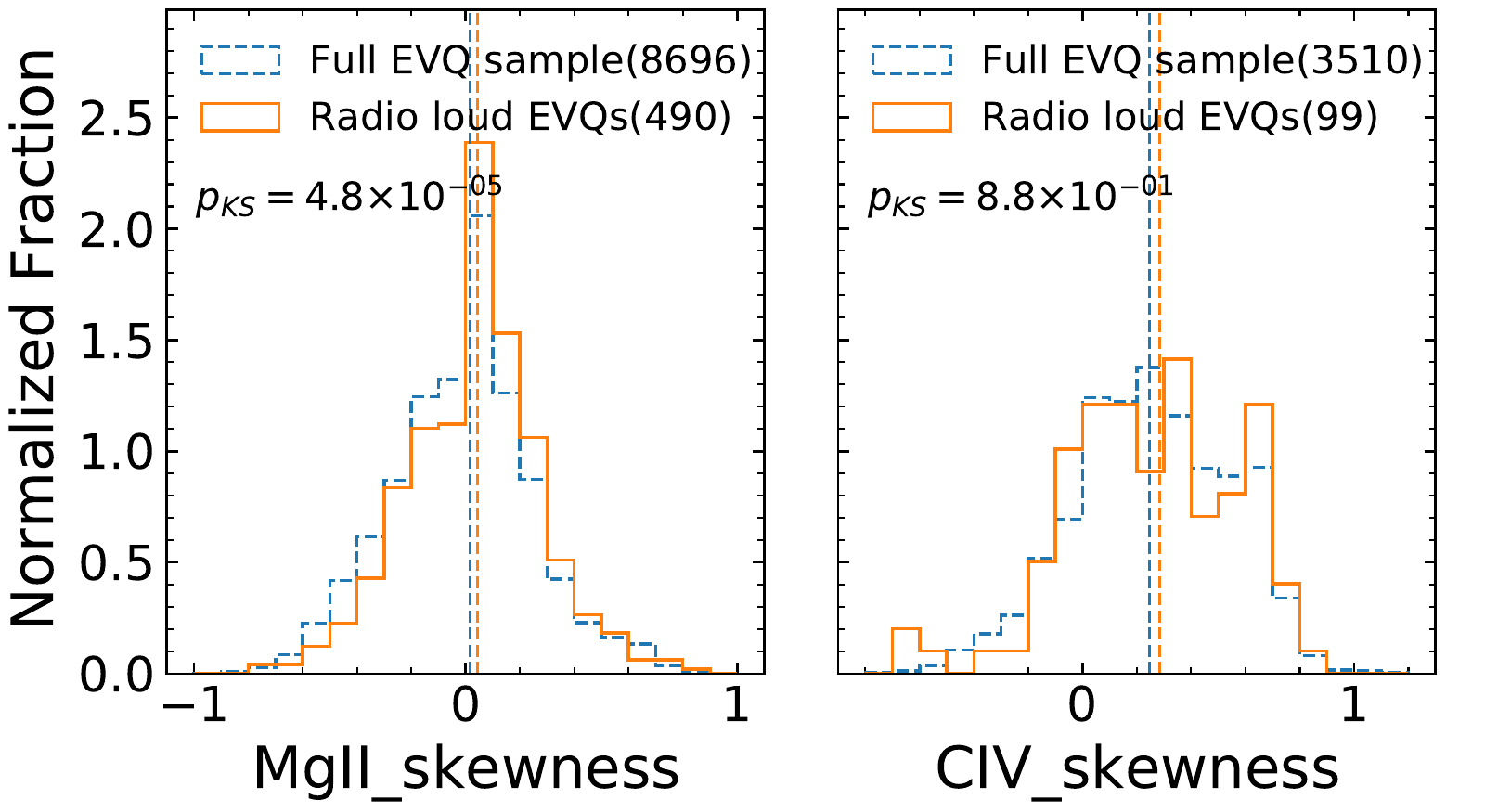}
   \caption{The distribution of the line skewness of radio loud EVQs vs the full EVQ sample. Numbers on the legend represent the sample size and the distribution are normalized. The median values for the samples are plotted as the vertical dashed lines. The K-S test results are also shown.\label{fig:radioskewness}}
\end{figure}

It's interesting to note that radio loud AGNs and blazars tend to show excess of redshifted very broad components of broad emission lines, particularly in high ionization line \CIV\ \citep{Punsly2020}, and the \CIV\ red wing luminosity excess was found to correlate with radio loudness (the spectral index from 10 GHz to 1350\AA, \citealt{Punsly2010}).
Such redshifted very broad component could be produced by gas lying deep within the gravitational potential of the central SMBH, and for the nearly face-on orientation in blazars, the gravitational redshift could be comparatively large \citep{Punsly2020}.
The redward excess is somehow similar to what we find in the \CIV\ profile in EVQs. 
However, limiting our study to radio quiet quasars (quasars with $f_{6cm}/f_{2500} >10$ based on FIRST detections, assuming a radio spectral index of $\alpha = -0.5$, are defined as radio loud, \citealt{Jiang2007}) does not alter the results in this work. 
Furthermore, the radio loud fraction in our EVQ sample ($\sim5.06\pm0.18\%$) is also comparable to that in the control samples ($\sim3.95\pm0.16\%$)\footnote{Two fractions are slightly different likely because the photometric selection of radio detected and non-detected quasar candidates for SDSS spectroscopic observations were not uniform \citep{Richards2002b}}. 
Thus the discoveries in this work are not due to a small fraction of radio loud quasars in our samples, but represent the properties of the general population of EVQs. 

In Fig. \ref{fig:radioskewness} we further plot the \MgII\ and \CIV\ line skewness of radio loud EVQs, compared with the full sample. The median skewness parameters and the K-S test indicate that radio loud EVQs do show (but slightly) more redward skewness (statistically marginal for \CIV). 

\cite{Punsly2020} also find that blazars with lower Eddington ratio tend to show strong redward asymmetry which can be explained as the Eddington ratio influenced the distance of the most efficient BLR to the center black hole.
However, in this work since the control samples were selected to have matched Eddington ratios, thus the stronger very broad component and more redward asymmetric \CIV\ line profile we found in EVQs can't not be attributed to lower Eddington ratios.

\subsection{\CIV\ Systematical Blueshift} \label{subsec:blueshift}

\begin{figure}[tb!]
   \includegraphics[width=.48\textwidth]{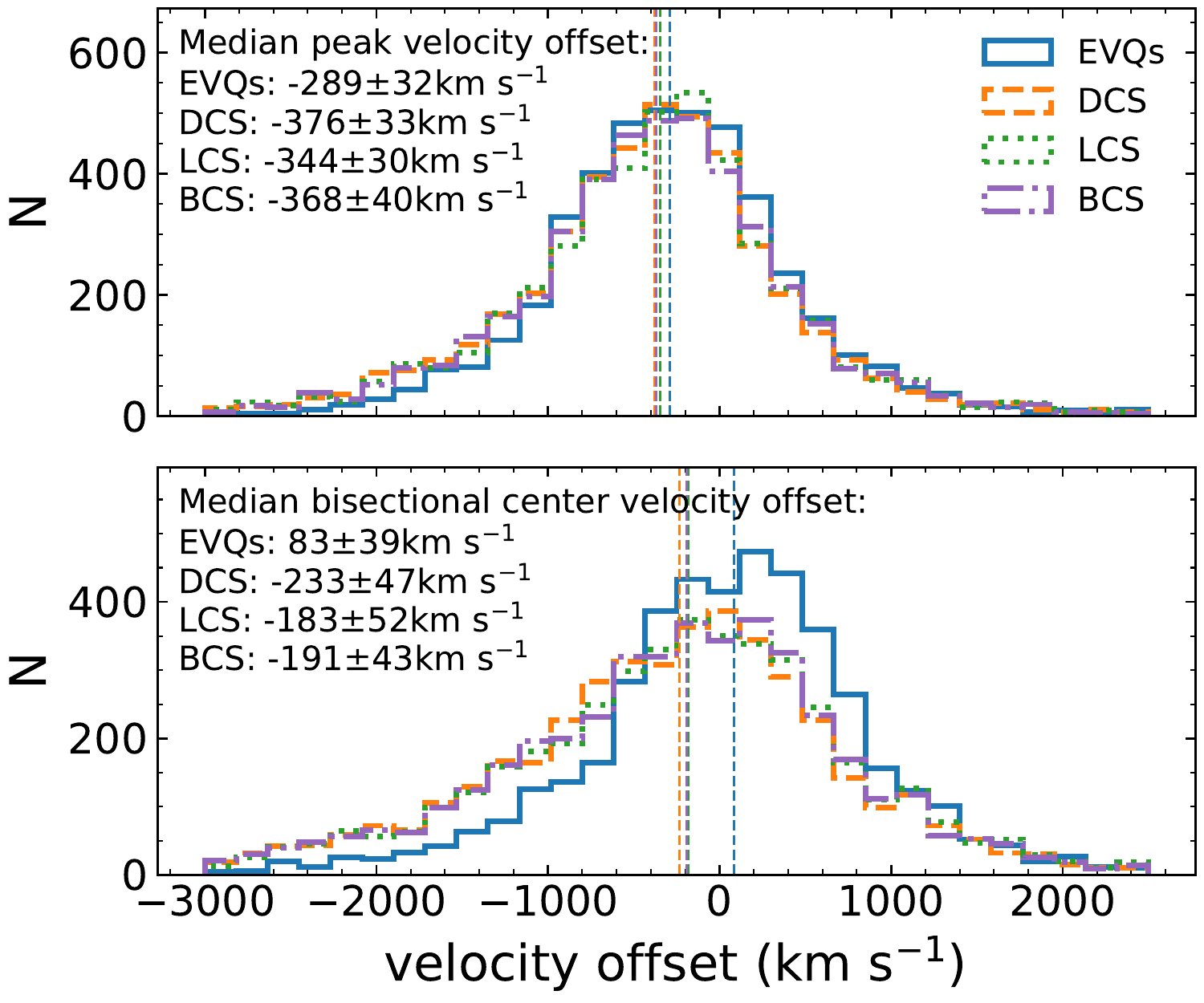}
   \caption{The \CIV\ line velocity offset (top: line peak; bottom: bisectional line center) with respect to the systematical redshift determined by \MgII\ line peak. The median values are marked with vertical lines and presented in the plot. 
   \label{fig:blueshift}}
\end{figure}

It has been widely reported that the high-ionization BELs in luminous quasars often significantly shifted blueward with respect to low-ionization lines \citep[e.g., ][]{Gaskell1982, Wilkes1986, Corbin1990,Sulentic2000, Sulentic2007, Richards2002a, Richards2011, Baskin2005a, Shen2008a, Wang2011, Denney2012, Shen2012, Coatman2016, Coatman2017, Sun2018d}.
Such blueshift is commonly contributed to accretion-disk winds \citep[e.g., ][]{Richards2011, Denney2012}.

We test whether the \CIV\ blueshift in EVQs differs from that in the control samples, using quasars with redshift 1.5 $<$ z $<$ 2.25 for which the low ionization \MgII\ line is available to derive the systematical redshift. We measure the velocity offset of \CIV\ line in each quasar, for both the line peak and the bisectional line center ($\lambda_b$ that bisect the total line flux, \citealt{Sun2018d}), 
with respect to the systematical redshift determined from the \MgII\ line peak. 
We find that while both EVQs and their control samples show clear \CIV\ blueshift on average (with negative median values of the offset), the median peak blueshift is marginally ($<$ 2$\sigma$) weaker in EVQs (see Fig. \ref{fig:blueshift}).
Note the K-S test shows that the distribution of \CIV\ line peak velocity offset in EVQs is different from that of the control sample DCS with a p-value of $6\times 10^{-10}$. This indicates that, compared with the median line peak blueshift, the line peak velocity offset distribution better reveals the difference between EVQs and the control sample.

We further find the \CIV\ bisectional line center of EVQs shows a redward-skewed distribution (with a positive median velocity offset), clearly different from their control samples (with negative offsets), and the K-S test yields a p-value of $5\times10^{-65}$ between EVQs and DCS. That is, because of the more redward-skewed line profile of \CIV\ in EVQs, though the line peak is blueshifted, the bisectional line center is contrarily redshifted. 
This pattern is consistent with what we have revealed from the stacked line profile (see \S\ref{subsec:profile}). The difference in the line profile may also partially account for the weaker \CIV\ peak blueshift in EVQs.
We note that the series weak emission lines (He\,\textsc{ii} $\lambda$1640, O\,\textsc{iii}] $\lambda$1663 and He\,\textsc{ii} $\lambda$1671) reside in the red wing of \CIV\ could also affect the bisectional line center.
However, considering that those lines are relatively weak and submerged in the flux of \CIV, it is very hard to measured those lines precisely with current low SNR spectra.
Nevertheless, given their line centers which are $\sim 100 \AA$ away from that of \CIV, they are not likely to affect the peak of \CIV\ nor to contribute the red excess of \CIV.

It has been theoretically proposed that the disc wind in quasars could be driven by either radiation pressure (continuum and/or UV line) or magneto-centrifugal forces, or some combination thereof \citep{Murray1995,deKool1995,Proga2000}. As EVQs have stronger emission line compared with their control samples with matched redshift, UV monochromatic luminosity and SMBH mass, they may have relatively stronger ionization continuum (see \S\ref{subsec:largerew}), thus stronger continuum radiation pressure is expected. 
The fact that the \CIV\ line peak blueshift in EVQs is weaker than that in the control samples disfavors the scenario that the wind acceleration is dominated by continuum radiation pressure.

\section{Conclusions} \label{sec:conclusion}
We have built a sample of 14,012 EVQs using the combined SDSS and PS1 light curves with a time span of $4\sim 15$ years and sorted them into different states according to the deviation of the spectra luminosity of each EVQ from its mean photometric luminosity.
Through comparing the EVQ samples in various states and well-defined control samples with matched redshift, luminosity, and SMBH mass, we derive the following main findings:

\begin{enumerate}
  \item The ``bluer-when-brighter'' pattern commonly seen in AGNs and quasars is clearly and similarly presented in EVQs (see Fig. \ref{fig:compareconti}). This finding suggests that the extreme variability might be due to the same mechanism as common variability. 
  \item We see significantly higher line EWs of broad \MgII\ and \CIV\ in dim states compared with those in brighter states. The trend is both qualitatively and quantitatively similar to the intrinsic Baldwin effect reported in literature in individual AGNs (see Fig. \ref{fig:ibeff}). However, it is puzzling that the broad \Hb\ line EW in EVQs shows no dependence on brightness state.
  \item We find that the EVQs have systematically greater EW, compared with control samples, in both broad (\Hb, \MgII\ and \CIV) and narrow (\OIII) lines (see Fig. \ref{fig:ewstack}). The EW excess of \MgII\ is the most prominent, reaching $\sim 47\%$. Such phenomenon could be related to the strong disc fluctuation/turbulence in EVQs, which may produce harder ionizing spectra and/or higher coverage of BLR.
  \item We find EVQs show subtle excess in the very broad line component, compared to their control samples (see Fig. \ref{fig:line_flux_norm}). This is likely because the stronger disc turbulence associated with the extreme variability in EVQs could launch relatively more gas from the inner disc into the very broad line region.
  \item EVQs show weaker \CIV\ line peak blueshift with respect to the systematic redshift derived from the \MgII\ line, compared with the control samples. While the blueshifted core of the \CIV\ line might comes from an outflowing intermediate line region,  
  the relatively stronger redshifted very broad component in EVQs makes the \CIV\ line more redward skewed in EVQs (see Fig. \ref{fig:civskewness}) compared with the control samples.
\end{enumerate}

In total, 2,341 of our EVQs have multi-epoch SDSS spectra, including 62 with at least 6 spectra. This provides us a good chance to explore the spectral variability of continuum and emission lines in a large sample of individual EVQs, which will be presented in a future work of this series.

The work is supported by National Natural Science Foundation of China (grants No. 11421303, 11873045, 11890693 \& 12033006) and CAS Frontier Science Key Research Program (QYZDJ-SSW-SLH006).
H.G. acknowledges the support from the NSF grant AST-1907290.
The authors thank Yue Zhao for useful comments and discussion.
This work has made use of SDSS photometric and spectroscopic data.
Funding for the Sloan Digital Sky Survey IV has been provided by the Alfred P. Sloan Foundation, the U.S. Department of Energy Office of Science, and the Participating Institutions. SDSS-IV acknowledges
support and resources from the Center for High-Performance Computing at
the University of Utah. The SDSS web site is www.sdss.org.
SDSS-IV is managed by the Astrophysical Research Consortium for the 
Participating Institutions of the SDSS Collaboration including the 
Brazilian Participation Group, the Carnegie Institution for Science, 
Carnegie Mellon University, the Chilean Participation Group, the French Participation Group, Harvard-Smithsonian Center for Astrophysics, 
Instituto de Astrof\'isica de Canarias, The Johns Hopkins University, 
Kavli Institute for the Physics and Mathematics of the Universe (IPMU) / 
University of Tokyo, the Korean Participation Group, Lawrence Berkeley National Laboratory, 
Leibniz Institut f\"ur Astrophysik Potsdam (AIP),  
Max-Planck-Institut f\"ur Astronomie (MPIA Heidelberg), 
Max-Planck-Institut f\"ur Astrophysik (MPA Garching), 
Max-Planck-Institut f\"ur Extraterrestrische Physik (MPE), 
National Astronomical Observatories of China, New Mexico State University, 
New York University, University of Notre Dame, 
Observat\'ario Nacional / MCTI, The Ohio State University, 
Pennsylvania State University, Shanghai Astronomical Observatory, 
United Kingdom Participation Group,
Universidad Nacional Aut\'onoma de M\'exico, University of Arizona, 
University of Colorado Boulder, University of Oxford, University of Portsmouth, 
University of Utah, University of Virginia, University of Washington, University of Wisconsin, 
Vanderbilt University, and Yale University.
This work has made use of PS1 photometric data.
The Pan-STARRS1 Surveys (PS1) and the PS1 public science archive have been made possible through contributions by the Institute for Astronomy, the University of Hawaii, the Pan-STARRS Project Office, the Max-Planck Society and its participating institutes, the Max Planck Institute for Astronomy, Heidelberg and the Max Planck Institute for Extraterrestrial Physics, Garching, The Johns Hopkins University, Durham University, the University of Edinburgh, the Queen's University Belfast, the Harvard-Smithsonian Center for Astrophysics, the Las Cumbres Observatory Global Telescope Network Incorporated, the National Central University of Taiwan, the Space Telescope Science Institute, the National Aeronautics and Space Administration under Grant No. NNX08AR22G issued through the Planetary Science Division of the NASA Science Mission Directorate, the National Science Foundation Grant No. AST-1238877, the University of Maryland, Eotvos Lorand University (ELTE), the Los Alamos National Laboratory, and the Gordon and Betty Moore Foundation.

\vspace{5mm}

\facilities{SDSS, PS1}

\software{PyQSOFit \citep{2018ascl.soft09008G}, NumPy \citep{numpy}, SciPy \citep{Virtanen_2020}, Matplotlib \citep{Matplotlib}, Astropy \citep{astropy}, Kapteyn \citep{KapteynPackage}, pandas \citep{McKinney_2010, McKinney_2011}, ds9 \citep{2003ASPC..295..489J}}

\restartappendixnumbering
\appendix
\section{An alternative re-binning strategy} \label{sec:rebin}

\begin{figure}[tb!]
   \includegraphics[width=.48\textwidth]{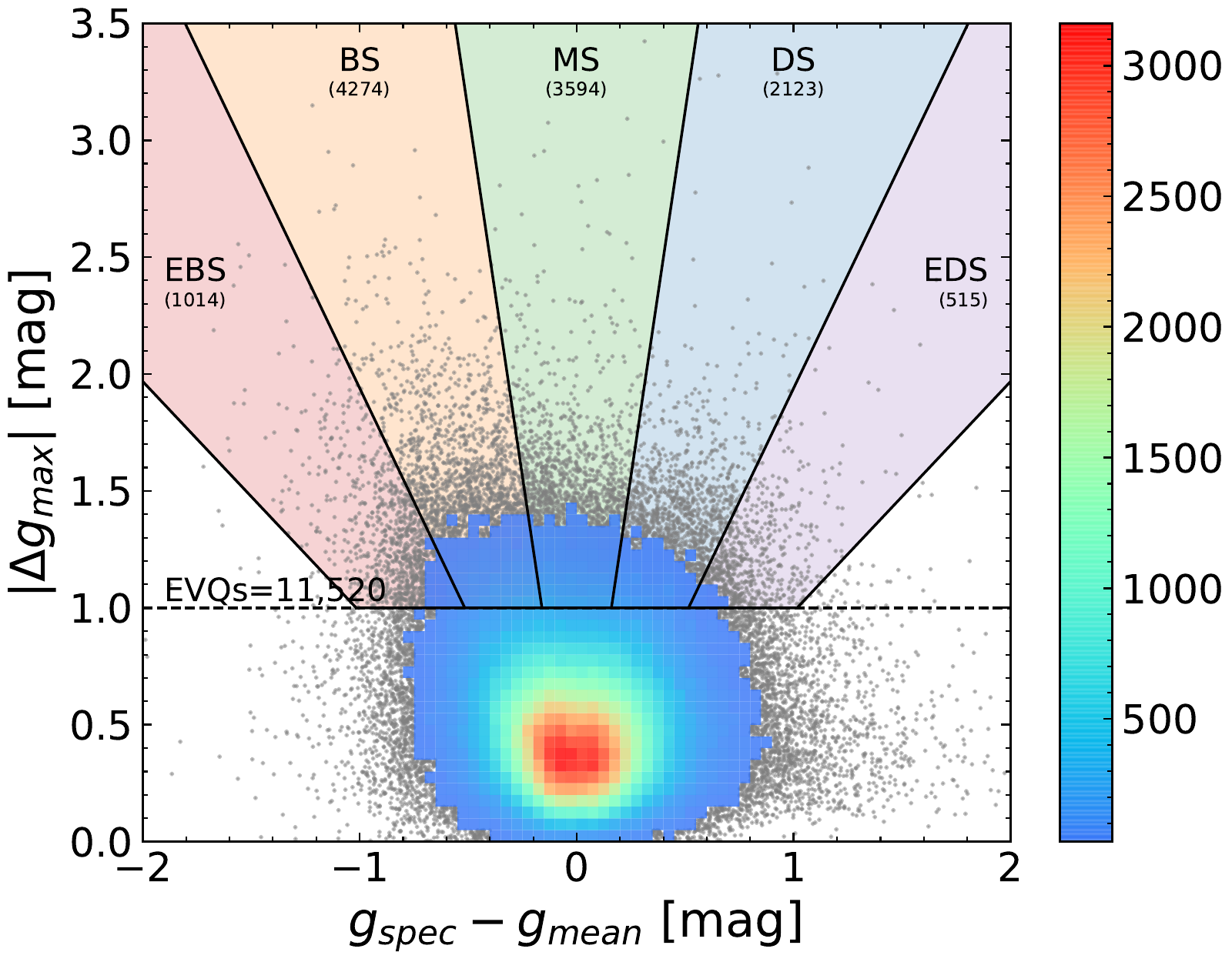}
   \caption{Similar to Fig. \ref{fig:evqsamp}, but using
   ($g_{\rm spec}-g_{\rm mean}$)/$\Delta g_{max}$ (instead of $g_{\rm spec}-g_{\rm mean}$ to re-bin the EVQ sample. 
   \label{figA:evqsamp}}
\end{figure}
\begin{figure}[tb!]
    \includegraphics[width=.48\textwidth]{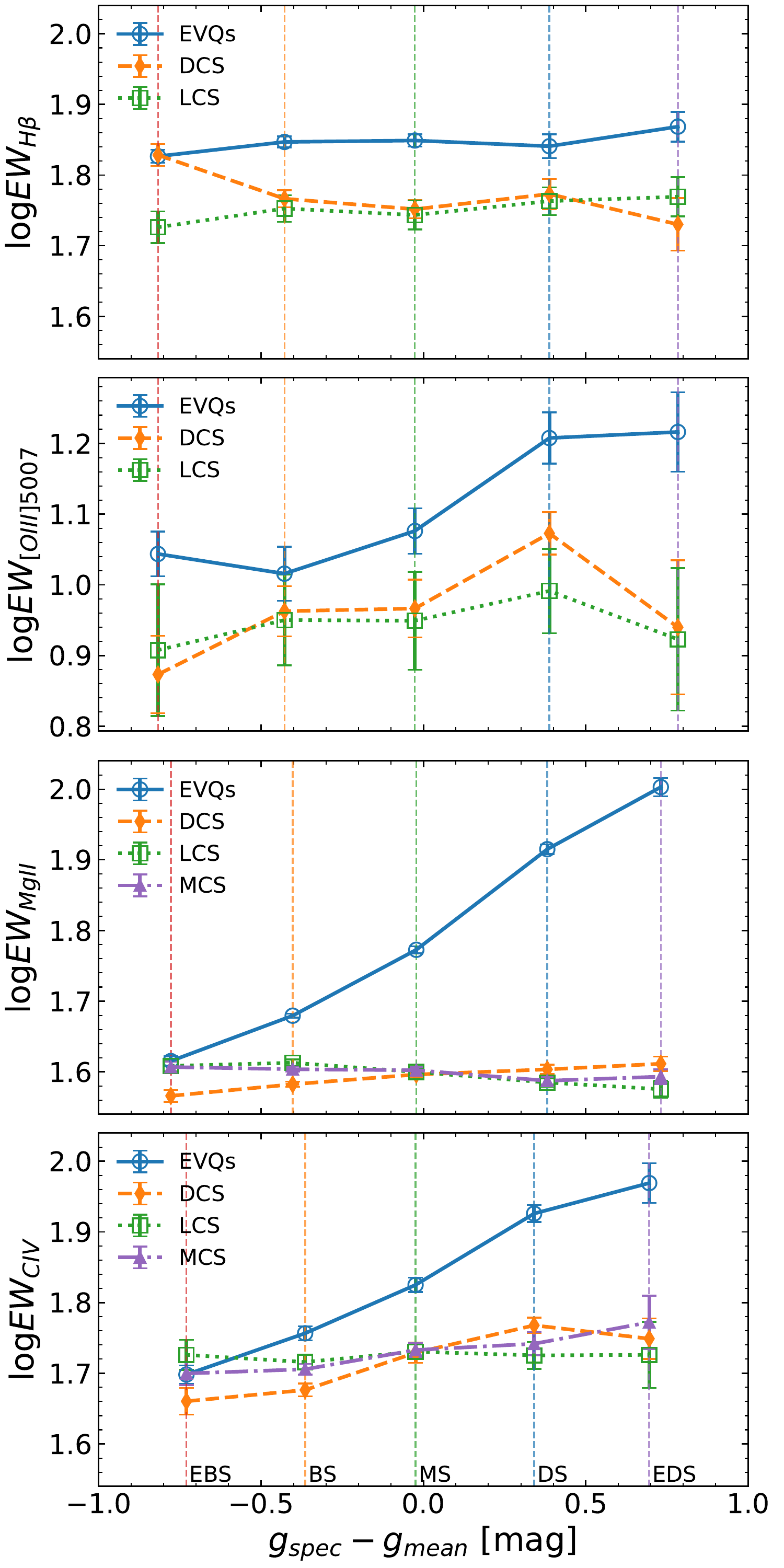}
    \caption{Similar to Fig. \ref{fig:ewstack} but the EVQs were divided into five classes using the alternative strategy demonstrated in Fig. \ref{figA:evqsamp}. 
    \label{figA:ewstack}}
\end{figure}
The underlying population of EVQs from extremely dim state to extremely bright state is continuous without gaps.
We divide the EVQs into five classes according to $g_{\rm spec}-g_{\rm mean}$ (see \S\ref{subsec:EVQ_sel}).
Alternatively, one may choose (${g_{spec}-g_{mean}})/\Delta g_{max}$ to define the spectra states (see Fig. \ref{figA:evqsamp}).
The new re-binning strategy however does not alter the results presented in this work. For instance,  
see Fig. \ref{figA:ewstack} for an updated version of Fig. \ref{fig:ewstack} with this new strategy.

\bibliography{ref}{}
\bibliographystyle{aasjournal}

\end{document}